\documentclass[conference]{IEEEtran}

\usepackage{tikz}
\usepackage{amsmath,amssymb}

\usepackage{filecontents}

\usepackage{tabularx}

\usepackage{graphicx}
\usepackage{xcolor}
\usepackage{amsthm}
\usepackage{booktabs}
\usepackage{float}
\usepackage{bm}
\usepackage{xurl}
\usepackage{epstopdf}

\usepackage{grffile}

\usepackage[font=small,skip=0pt]{caption}

\usepackage[T1]{fontenc}
\usepackage[utf8]{inputenc}
\usepackage{ae,aecompl}
\usepackage{wrapfig}

\usepackage{hyperref}

\theoremstyle{definition}
\newtheorem{definition}{Definition}[section]
\DeclareMathOperator{\E}{\mathbb{E}}

\newcommand{\point}[1]{\par\smallskip\noindent\textbf{#1.}}
\newcommand{\addr}[2][\small]{{#1\href{https://etherscan.io/address/#2}{\texttt{#2}}}}

\newcommand\todaysdate{April~15th,~2020}
\newcommand\lendingmarketshare{76\% }
\newcommand\makersharelending{65\% }
\newcommand\deficapitalization{702m USD }
\newcommand\definitionrelevance{93\% }

\newcommand\makercapitalization{342.9m }
\newcommand\compoundcapitalization{91.6m }
\newcommand\aavecapitalization{36.4m }
\newcommand\uniswapcapitalization{35.7m }
\newcommand\bancorcapitalization{7.2m } 
\newcommand\kybercapitalization{3.9m }
\newcommand\synthetixcapitalization{101.9m }
\newcommand\nexuscapitalization{2.7m }
\newcommand\erasurecapitalization{1.2m }\newcommand\lightningcapitalization{6.5m }
\newcommand\connextcapitalization{12.1k }
\newcommand\tokensetscapitalization{9m }
\newcommand\wbtccapitalization{7.3m }
\newcommand\meloncapitalization{221.9k }

\PassOptionsToPackage{hyphens}{url}\usepackage{hyperref}

\usepackage{enumitem}
\setitemize{noitemsep,topsep=0pt,parsep=0pt,partopsep=0pt}
\setdescription{noitemsep,topsep=0pt,parsep=0pt,partopsep=0pt}
\setenumerate{topsep=0pt,itemsep=-1ex,partopsep=1ex,parsep=1ex}

\begin{document}


\date{}



\title{\Large \bf The Decentralized Financial Crisis}


\author{
    \IEEEauthorblockN{
    Lewis Gudgeon\IEEEauthorrefmark{1}, 
    Daniel Perez\IEEEauthorrefmark{1}, 
    Dominik Harz\IEEEauthorrefmark{1}, 
    Benjamin Livshits\IEEEauthorrefmark{1} and 
    Arthur Gervais\IEEEauthorrefmark{1}}
    \IEEEauthorblockA{
    \IEEEauthorrefmark{1}
    Department of Computing, Imperial College London}
    \thanks{AAAA}
}

\maketitle

\thispagestyle{plain}
\pagestyle{plain}

\begin{abstract}

The Global Financial Crisis of~2008, caused by the accumulation of excessive financial risk, inspired Satoshi Nakamoto to create Bitcoin. 
Now, more than ten years later, Decentralized Finance~(DeFi), a peer-to-peer financial paradigm which leverages blockchain-based smart contracts to ensure its integrity and security, contains over \deficapitalization of capital as of \todaysdate. 
As this ecosystem develops, it is at risk of the very sort of financial meltdown it is supposed to be preventing. 
In this paper we explore how design weaknesses and price fluctuations in DeFi protocols could lead to a DeFi crisis.
We focus on DeFi lending protocols as they currently constitute most of the DeFi ecosystem with a~\lendingmarketshare market share by capital as of \todaysdate. 

First, we demonstrate the feasibility of attacking Maker's governance design to take full control of the protocol, the largest DeFi protocol by market share, which would have allowed the theft of 0.5bn USD of collateral and the minting of an unlimited supply of DAI tokens.
In doing so, we present a novel strategy utilizing so-called \textit{flash loans} that would have in principle allowed the execution of the governance attack in just \emph{two transactions} and without the need to lock any assets.
Approximately two weeks after we disclosed the attack details, Maker modified the governance parameters mitigating the attack vectors.
Second, we turn to a central component of financial risk in DeFi lending protocols.
Inspired by stress-testing as performed by central banks, we develop a stress-testing framework for a stylized DeFi lending protocol, focusing our attention on the impact of a \textit{drying-up} of liquidity on protocol solvency.
Based on our parameters, we find that with sufficiently illiquidity a lending protocol with a total debt of 400m USD could become undercollateralized within 19 days. 
\end{abstract}


\section{Introduction}
\label{sec:introduction}
Blockchain technology emerged as a response to the Financial Crisis of 2007--8~\cite{nakamoto2008bitcoin}.\footnote{The first Bitcoin block famously outlines: ``The Times 03/Jan/2009, Chancellor on brink of second bailout for banks''.}
The perception that banks had misbehaved resulted in a deterioration of trust in the traditional financial sector~\cite{earle2009trust}.
The causes of the crisis were several, but arguably chief among them was a lack of transparency regarding the amount of risk major banks were accumulating.
When Lehman Brothers filed for bankruptcy, it had debts of~613bn USD, bond debt of 155bn USD and assets of~639bn USD~\cite{bbc}. 
Central to its bankruptcy was its exposure to subprime (i.e., bad quality) mortgages.
This exposure was compounded by the fact that the bank had a leverage ratio\footnote{Defined as $\frac{\text{total assets}}{\text{equity}}$; the total assets were more than~30 times larger than what shareholders owned, indicating substantial debt.} of~30.7x in~2007~\cite{secinfo}.

From their inception, blockchain-based cryptocurrencies sought to provide a remedy to such crises: facilitating financial transactions without reliance on trusted intermediaries,  shifting the power, and therefore, the ability to cause crisis through the construction of opaque and complex financial instruments, away from banks and financial institutions~\cite{nakamoto2008bitcoin}.
Ten years later, a complex financial architecture---the architecture of Decentralized Finance (DeFi)---is gradually emerging on top of existing blockchain platforms.
Components in this architecture include those that pertain to lending, decentralized exchange of assets, and markets for derivatives (cf.\ Appendix~\ref{appendix:existing-defi-protocols} Table~\ref{tab:defiprojects})~\cite{makerdao,compoundfinance, synthetix, uniswap, dydx, instadapp}.

DeFi architectures for lending require agents to post \textit{security deposits} to fully compensate counter-parties for the disappearance of the agent.
We assume that when an economically rational agent faces a choice between the repayment of a debt or the loss of collateral, given the absence of reputation tracking---on account of agent pseudonymity and the possibility of an agent using multiple addresses---the agent will choose the least costly option.
Security deposits serve to guard against~(i) \textit{misbehavior} of agents, where the action that would maximize individual utility does not maximize social welfare, and (ii) external events, such as large exogenous drops in the value of a particular cryptocurrency ~\cite{harz2019balance}.
Of all DeFi protocols, those with the most locked capital are for lending.
As of \todaysdate, the largest protocol by capitalization, Maker~\cite{makerdao}, has c.~\makersharelending of all capital locked in DeFi, corresponding to~\makercapitalization USD~\cite{defipulse}.

\textit{Governance} is another crucial facet of DeFi protocols and we observe differing degrees of governance decentralization. 
For example, Maker uses its own token (MKR) to allow holders to vote on a contract that implements the governance rules. 
In contrast, Compound~\cite{compoundfinance}, the third largest protocol by market share, is centrally governed and a single account can shut down the system in case of a failure.
Moreover, as in traditional finance, these protocols do not exist in isolation.
Assets that are created in Maker, for example, can be used as collateral in other protocols such as Compound, dYdX~\cite{dydx}, or in liquidity pools on Uniswap~\cite{uniswap}. 
Indeed, the composability of DeFi~---~the ability to build a complex, multi-component financial system on top of crypto-assets~---~is a defining property of open finance~\cite{money-legos}. 
However, if the underlying collateral assets fail, all connected protocols will be affected as well: there is the possibility of financial contagion.

\textbf{This paper.}
We focus on DeFi \textit{lending} protocols, which constitute c.~\lendingmarketshare of the DeFi market in capital terms.
We consider two distinct but interconnected aspects of the attack and risk surface for collateral-based Defi lending protocols: (i) attacks on the governance mechanism and (ii) the economic security of such protocols in ``black swan''~\cite{taleb2007black} financial scenarios. 

In relation to attacks on the governance mechanism, we examine an attack on Maker~\cite{makerdao} which consists of an adversary amassing enough capital to seize full control of the funds within the protocol.
Herein, we further consider two distinct attack strategies.
We engaged in a process of responsible disclosure with Maker, which we detail, who since modified their governance parameters to mitigate the two attack strategies we present. 
The first attack strategy, crowdfunding, inspired by~\cite{maker-governance-attack}, covertly executed, was feasible within two blockchain blocks and required the attacker to lock c.~27.5m USD of collateral.
This would have enabled an attacker to steal all 0.5bn USD of locked collateral in the protocol and mint an unlimited supply of DAI tokens.
The second, novel, attack strategy utilized so-called flash loans and allows an adversary to amass the Maker collateral within a single transactions.  
This attack did not require locking of collateral and only required a few US dollars to pay for gas fees.

With respect to the economic security of collateralized DeFi lending protocols, with reference to our stylized model, we present the possibility of a DeFi lending protocol becoming undercollateralized (or insolvent)~---~where security deposits become smaller than the issued debt~---~as the result of a drying up of liquidity.
Assuming rational economic agents, in such an under-collateralization event, the borrower would default on their debts, since the amount they have borrowed has become worth more than the amount they escrowed.
Starting from formal definitions of the economic security constraints for DeFi lending platforms, we then use Monte Carlo simulation to stress-test their financial robustness.
We submit that such stress-testing constitutes an important approach to bounding the economic security of DeFi lending protocols when formal security proofs are not obtainable and their security primarily depends on economic properties.  
We find that for plausible parameter ranges, a DeFi lending system could find itself undercollateralized.
To the extent that other DeFi protocols allow agents to lend or trade the undercollateralized asset, financial contagion---where an economic shock spreads to other protocols---would be expected to result.

\point{Contributions}
\begin{itemize}

    \item {\bf Governance attack on Maker (Section~\ref{sec:governance-attack}).}
    With specific focus on the largest DeFi project by market share, Maker, we show how, prior to Maker implementing a parameter change, it was feasible to successfully steal the funds locked in the protocol covertly and within two blocks or within two transactions. 
    By exploiting recent flash loan pool contracts, we show how an attacker with no capital (besides gas fees), would have been able to execute such an attack, if the flash loan pools would provide sufficient liquidity (which they did not at the time of writing).

    \item {\bf Formal modeling of DeFi lending protocols (Section~\ref{sec:defilending}).}
    We provide definitions for economically-resilient DeFi lending protocols, introducing overcollateralization, liquidity, and counter-party risk as formal constraints.
    These definitions serve to formalize financial risk constraints for more than \definitionrelevance of the funds locked in DeFi lending protocols as of \todaysdate~\cite{defipulse}.
    
    \item {\bf Financial stress-testing (Section~\ref{sec:stress-test}).}
    We develop a methodology to quantitatively stress-test a DeFi protocol with respect to its financial robustness, inspired by risk assessments performed by central banks in traditional financial systems.
    We simulate a price crash event with our stress-test methodology to a stylized DeFi lending protocol that resembles the largest DeFi lending protocols to-date, by volume: Maker, Compound, Aave and dYdX. 
    We find, for plausible parameter ranges, that a DeFi lending protocol could become undercollateralized within 19 days.
    
\end{itemize}

\section{Overview of Decentralized Finance}
\label{sec:defi}

This section provides a definition of DeFi, describes the composability property of DeFi protocols, and states our assumptions regarding the underlying blockchain. 

DeFi is an emergent field, with over \deficapitalization of total value locked in DeFi protocols as of \todaysdate~\cite{defipulse}. 
Table~\ref{tab:defiprojects} in Appendix~\ref{appendix:existing-defi-protocols} presents a categorization of DeFi protocols, providing the three largest by locked USD in each case. 
We observe that Maker dominates the DeFi projects with a capitalization of over \makercapitalization USD. 
DeFi protocols mostly emerge for uses such as lending, decentralized exchange, and derivatives. 
We define DeFi as follows.

\begin{definition}[Decentralized Finance (DeFi)]
a peer-to-peer financial system, which leverages distributed ledger-based smart contracts to ensure its integrity and security.
\end{definition}

In this paper we make the assumption that agents are rational, that is, are agents who seek to maximize their expected utility.
We assume that agents are pseudonymous.

\subsection{DeFi Composability}

DeFi protocols do not exist in isolation.
Their open nature allows developers to create new protocols by composing existing protocols together.
Some compare this approach with ``Money Lego''~\cite{money-legos}.
As such, assets created through lending in one protocol can be reused as collateral in other protocols in any kind of fashion.
This creates a complex and intertwined system of assets and debt obligations.
Moreover, a failure of a protocol that serves as backing asset to other protocols has a cascading effect on others.
Indeed, a hallmark of financial crisis is that such events do not take place in isolation, but rather financial contagion takes place, where a shock affecting a few institutions spreads by contagion to the the rest of the financial sector, before affecting the larger economy~\cite{allen2000financial}.

\subsection{Blockchain Model}
\label{sec:blockchainmodel}

A DeFi protocol operates on top of a layer-one blockchain, which provides standard ledger functionality~\cite{badertscher2017bitcoin,badertscher2018ouroboros, david2018ouroboros,pass2017analysis}.
We assume that the underlying blockchain is able to provide finality~\cite{Castro1999, Miller2016a}, construed as a guarantee that once committed to the blockchain a transaction cannot be modified or reversed.
For this paper, we treat attacks on layer-one blockchains as orthogonal and as such we are not concerned with them.\footnote{Future work could consider the possibility of combining layer-one attacks with smart-contract attacks to increase their probability of success. For example, see the successful attack on Fomo3D~\cite{fomo3d}.}

\section{Governance Attack on Maker}\label{sec:governance-attack}

In this section, we first present an attack on the governance mechanism of the Maker protocol~\cite{makerdao}.
We use a representation of the state of the Ethereum main network on February 7th and the Maker contract to simulate as realistically as possible how such an attack could take place.
While focusing on a specific protocol, we submit that such a governance attack is representative of a new element of the attack surface for DeFi protocols more generally.
Since we first analyzed this attack vector, the Maker protocol has been modified to mitigate this attack: we detail our interaction with the Maker team below.
Although the basic idea of the attack had been briefly presented in a blog post~\cite{maker-governance-attack}, the \textit{feasibility} of the attack has not been analyzed.
 
\subsection{Disclosure to Maker}
\label{sec:maker-interaction}
We engaged in a process of responsible disclosure with Maker, as detailed below. 
\begin{itemize}
    \item On February~7th,~2020 we reached out to the Maker team regarding our exposition of the feasibility of the governance attack.
    \item On February~14th the authors had a conference call with Maker, where we described our work. 
    We agreed to giving the Maker team sight of this paper prior to our publication of it; we subsequently sent a draft of the paper on February~17th. 
    \item On February~18th the authors further contacted Maker to describe how the use of flash loans increased the risk of the governance attack, offering a response window prior to publicizing this result within which Maker provided helpful feedback.
\end{itemize}
After this exchange, on February~21st Maker announced that the Governance Security Module had been activated with a delay period of 24 hours~\cite{maker-24-hour}, mitigating the vulnerability.

\subsection{Background and Threat Model}
The governance process relies on the MKR token, where participants have voting rights proportional to the amount of MKR tokens they lock within the voting system. 
MKR can be traded on exchanges~\cite{mkr-coinmarketcap}.

\point{Executive voting} 
Using executive voting, participants can elect an \emph{executive contract}, defining a set of rules to govern the system, by staking (i.e., locking-up) tokens on it. 
Executive voting is continuous, i.e., participants can change their vote at any time and a contract can be newly elected as soon as it obtains a majority of votes. 
The elected contract is the only entity allowed to manipulate funds locked as collateral. 
If a malicious contract were to be elected, it could steal all the funds locked as collateral.

\point{Defense mechanisms} Several defense mechanisms exist to protect executive voting.
The \emph{Governance Security Module} encapsulates the successfully elected contract for a certain period of time, after which the elected contract takes control of the system. 
At the time of first writing on February~7th, this period was set to zero~\cite{mkr-vote-failed}.
This has subsequently been increased to~24 hours~\cite{maker-gsm}, see Section~\ref{sec:maker-interaction}.
The~\emph{Emergency Shut Down}, which allows a set of participants holding a sufficient amount of MKR to halt the system.
However, this operation requires a constant pool of~50k MKR tokens, worth~27.5M USD as of February~7th.\footnote{We use the price of MKR on~2020-02-01, which was~550 USD.}

\point{Threat model}
We assume the existence of a rational adversary i.e., one who would only engage in the attack if the potential returns are higher than the costs.
In this attack, the costs are the amount of money that the adversary has to pay to have his contract elected as executive contract. The returns are the amount of money that the contract could steal or generate once it is elected.
There are two ways in which electing an adversarial executive contract can financially benefit the adversary. 
First, the contract can transfer all the ETH collateral to the adversary's address. 
Second, the contract can mint new DAI tokens and transfer them to the adversary. 
The DAI tokens can then be traded until the DAI price crashes and effectively destroy the Maker system.

As of February~7th there were c.~150k MKR tokens used for executive voting and the current executive contract had~76k MKR tokens staked. 
We observed that the staked amount changes relatively often and the amount of tokens staked on the elected contract often dropped below~50k MKR tokens (eq.~27.5M USD). 
As of February~7th, there were c.~470M USD worth of ETH locked as collateral of the DAI supply, which an executive contract can dispose of freely.
This shows that even before trading the DAI tokens, the attack would have been financially attractive.

\subsection{Crowdfunding and Flash Loans}
An adversary can choose between the two following strategies to amass the capital required for the governance attack.

\point{Crowdfunding}
Crowdfunding MKR tokens may allow users to lock their tokens in a contract and program the contract so that when the required amount of MKR tokens is reached, it stakes all its funds on a malicious executive contract.
This would allow multiple parties to collaborate trustlessly on such an attack, while keeping control of their funds and while being assured that they will be compensated for their participation in it.
\footnote{An (admittedly informal) poll on Twitter from late 2019 conducted by a user soon after this attack first appeared shows that several participants would be interested in such crowdfunding. 
See Fig.~\ref{fig:mkr-crowdfunding-tweet} in Appendix~\ref{appendix:governance-attack-on-maker}.}

\begin{figure}[tb]
  \centering
  \includegraphics[width=\columnwidth]{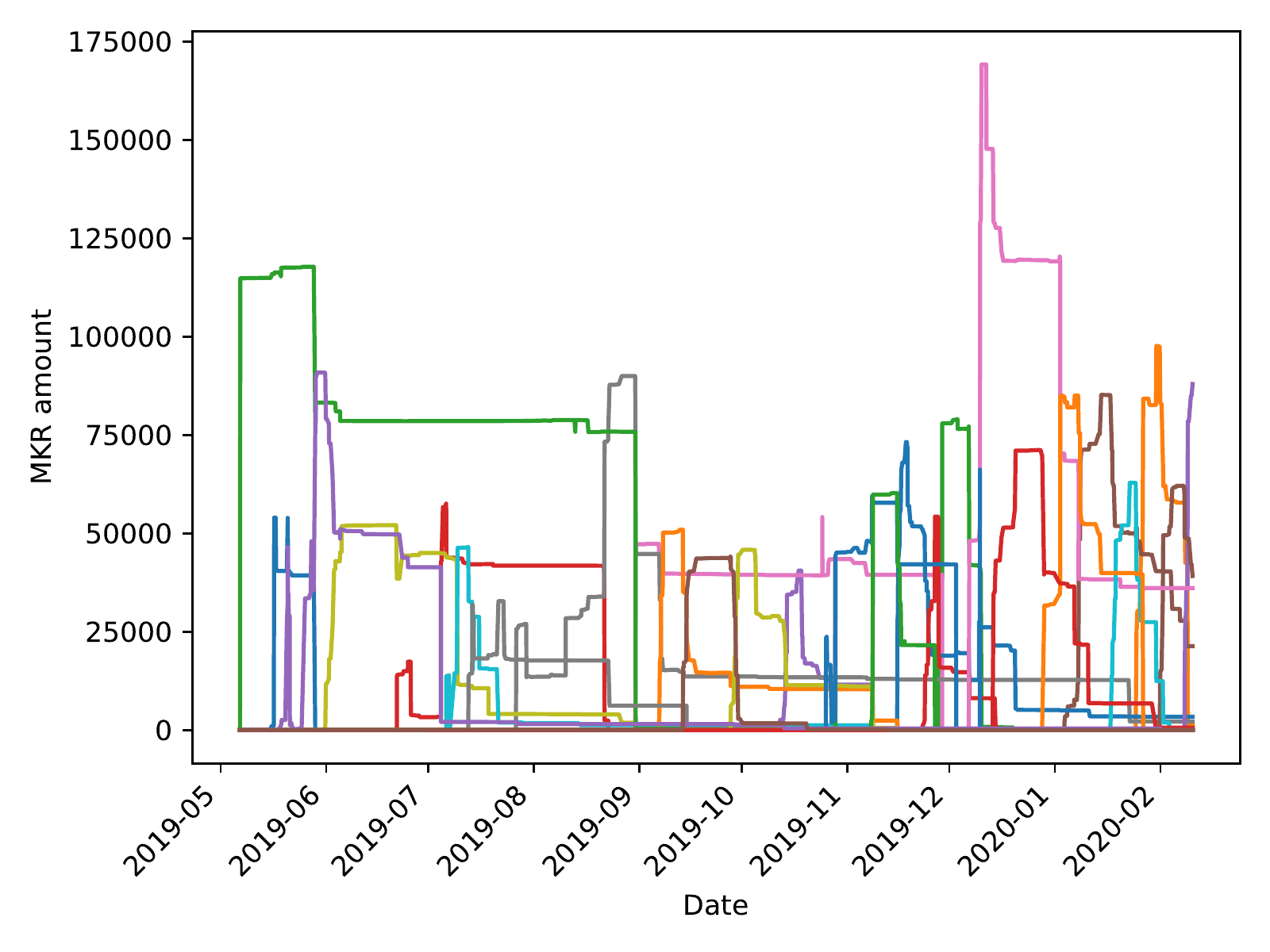}
  \caption{Evolution of the amount of MKR tokens staked on different executive candidate contracts. We observe that at times the MKR amount of the executive contract dropped below 50k MKR.}
  \label{fig:mkr-staked}
\end{figure}

\point{Liquidity pools and flash loans}
A shortcoming of the crowdfunding attack is the required coordination effort between the participants and the likely alerting of benevolent MKR members. 
Instead, an attack could use liquidity pools offering \textit{flash loans}~\cite{aave}. 
A flash loan is a \emph{non-collateralized} loan that is valid within one transaction only. 
In the Ethereum Virtual Machine (EVM), a transaction can be reverted entirely if a condition in one part of the transaction is not fulfilled.
A flash loan then operates as follows: a party creates a smart contract that (i) takes out the loan, (ii) executes some actions, and (iii) pays back the loan and, depending on the platform, interest.

The interesting aspect for our purposes is that if in step (ii) the execution of the actions fails or step (iii) the repayment of the loan cannot be completed, the EVM treats this loan as it never took place. 
Hence, under the assumption there is enough liquidity available in protocols such as dYdX~\cite{dydx} and Aave~\cite{aave-pools}, an attacker could execute the MKR governance attack in step (ii), and, if successful, repay the flash loan in step (iii). 
Since the flash loan requires no collateral, the capital lock up cost for the attacker is significantly reduced. 
If there is enough liquidity available in these pools, the attacker might even not have to lock any tokens.
Furthermore, the liquidity provider may have also profited from the execution of the attack, depending on in which protocol their tokens were locked.
For example, in Aave as of February~7th they would have received an interest rate of 0.09\% for each flash loan.

\subsection{Practical Attack Viability}

In this section, we use empirical data to show how such an attack could take place, and describe what the potential shortcomings could be.
We first analyze all the transactions received by Maker's governance contract of as February~7th: 
\addr{0x9ef05f7f6deb616fd37ac3c959a2ddd25a54e4f5}.
Since the deployment of this contract, in May~2019, there were~24 different contracts which have been elected as executive contract (cf.\ Fig.~\ref{fig:mkr-staked}).
When a contract is elected as the executive contract, the total amount of staked MKR is, for a short period of time, distributed almost equally  between the old and the new executive contract.\footnote{This was particularly visible at the end of November~2019~(80k MKR to 40k MKR) and in the middle of January~2020~(120k MKR to 45k MKR).}
This serves to reduce the amount of tokens required for the attack by more than 50\%.

One day after the first blog on this attack was published~\cite{maker-governance-attack}), there was a sharp increase in the MKR staked on the executive contract, rising from c.~75k to c.~160k MKR at the beginning of December~2019.
One token holder~\cite{maker-voting-distribution} in particular injected a large quantity of tokens---c.~66k MKR---potentially to help prevent an attack from occurring.
\footnote{It is unclear if the token holder was the Maker Foundation or some other party; in our discussion with Maker they stated they knew the identity of the token holder. 
The holder staked their tokens on the currently elected contract, making the attack more difficult to execute, before releasing the staked tokens approximately one month later. 
The token holder had more than sufficient tokens to execute the attack:  were they malicious, they could have stolen the funds.}

\subsection{The Attacks}


\begin{figure*}[tb]
    \centering
    \includegraphics[width=0.8\textwidth]{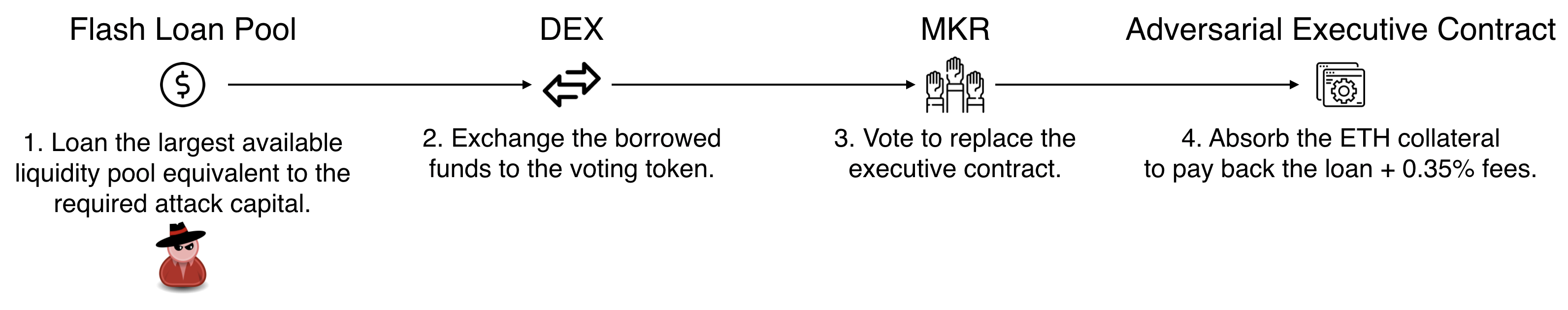}
    \caption{Example flash loan attack against Maker. All steps can be executed within one transaction, under the assumption that the flash loan pool and DEX have sufficient liquidity available. To execute the attack, the adversary would not need upfront capital, besides the gas fees (estimated to amount to c.~15 USD).}
    \label{fig:flash-loan-attack}
\end{figure*}

\point{The crowdfunding strategy}
We inspected the amount of MKR transferred between January~1st,~2020 and February~8th,~2020.
\footnote{See Appendix~\ref{appendix:governance-attack-on-maker}, Fig.~\ref{fig:mkr-traded-volume}.}
We find a mean MKR transaction volume of c.~9k MKR tokens per day, corroborated by e.g.~\cite{mkr-coinmarketcap}.
Given such volumes, an attacker accumulating 1k MKR tokens per day, for instance, would have sufficient tokens in less than 2 months.
However, accumulating all the tokens in a single account would likely attract attention. 
Indeed, from our discussions with the Maker team, the large MKR token holders seem to be known.

To be covert, an attacker could try to accumulate tokens to multiple accounts without perceptibly changing the distribution of MKR tokens.
On February~8th there were c.~5k accounts, holding a total of c.~272k MKR tokens.\footnote{Excluding the holders with a low balance (less than~1 MKR token), and a large balance (more than~5k MKR tokens).}
Given that the attack is possible with 50k MKR tokens, an adversary could spread their tokens across e.g. 100 accounts with an average of 500 tokens each.
However, one drawback of this approach is the requirement to vote from these 100 accounts.
Voting for a contract costs on average~69k gas. Given the gas limit per block on Ethereum is 10 million, filling half of a block with voting transactions would allow votes from $10\text{M} / 69\text{k} \approx 72$ contracts.
Doing so would be inexpensive~\cite{perez2019broken}, meaning that an attacker would have been able to easily perform the whole attack in two blocks.
In the second block, the attacker would finish voting for his malicious contract and execute the attack from the contract, which would leave only one block for anyone to react to the attack.

\point{The flash loan strategy}
Alternatively, to execute the governance attack without amassing tokens, the attacker could utilize liquidity pools to borrow the required tokens via a flash loan (e.g.\ via dYdX~\cite{dydx} or Aave~\cite{aave}).
\footnote{Aave is a protocol deployed on the Ethereum mainnet on January~8th,~2020, \url{https://etherscan.io/tx/0x4752f752f5262fb11733e0136033f7d53cdc90971441750f606cf1594a5fde4f}.} 
The attacker makes the following two transactions (cf.\ Fig.~\ref{fig:flash-loan-attack}).

\begin{description}
    \item[Transaction 1:] Deploy the malicious governance contract and deploy the attack contract executing the flash loan.
    \item[Transaction 2:] Call the attack contract deployed in Transaction 1 that executes the following steps.
    \begin{enumerate}
        \item Take out flash loan(s)(e.g. from Aave and dYdX) in the currency with the deepest markets for buying MKR tokens.
        As of February 7th this was ETH.
        \item Sell the ETH loan for 50k MKR tokens on decentralized exchange(s).
        \item Vote with the 50k MKR tokens to replace the current Maker governance contract with the malicious contract deployed in Transaction 1.
        \item Mint DAI into an account chosen by the attacker. 
        \item Take out enough ETH from the Maker system to repay the flash loan.
        \item Repay the flash loan with the required 0.09\% interest to Aave and repay the flash loan to dYdX with minimal (1 WEI) interest.
    \end{enumerate}
\end{description}


In a naive approach, we could utilize the exchange rate for ETH to MKR to obtain that an attacker requires 114,746 ETH to execute the attack.
However in practice, an attacker seeking to buy such a large quantity of MKR tokens would pay a greater price than this, forced to buying the tokens at the best remaining market price for each unit.
As of February~14th, an attacker sourcing the required 50k MKR tokens from three different DEXs---38k MKR from Kyber, 11,500 MKR from Uniswap, and~500 MKR from Switcheo---would need a total of~378,940 ETH, 3.3x that of the naive estimate.
\footnote{For current liquidity and rates see \url{https://dexindex.io/}.}
As of February~14th, the flash loan providers had insufficient pool liquidity: dYdX had c.~83,590 ETH and Aave had c.~13,670 ETH. 
However, on February~14th the ETH growth rate of Aave was~5.18\% per day.
Assuming the growth rate continued, it would have only taken \emph{66 days} until enough liquidity was available in Aave.

\subsection{Profitability Analysis}

\point{The crowdfunding strategy}
With the crowd funding strategy, the profits from the attack could be split equally between the funding parties.
The only cost are the~20 USD for including the transactions~\cite{perez2019broken}.
In return, the attackers could take away the 434,873 ETH in collateral in Maker plus 145m DAI, amounting to a net profit of 263m USD (as of February~7th).
Additionally, the attackers could mint unlimited new DAI and use this to buy other cryptocurrencies available at centralized and decentralized exchanges.

\point{The flash loan strategy}
Assuming dYdX's and Aave's liquidity pools had accumulated the required~378,940 ETH to execute the attack, we can calculate the profitability as follows.
The attacker obtains a total of~434,873 ETH in collateral from Maker as well as the~50k MKR tokens and the~22m DAI currently in circulation.
The attacker needs to repay the 378,940 ETH loan with minimal interest (1 WEI for dYdX and ~0.09\% for Aave~(265.82 ETH)).
Furthermore, the attacker needs to pay for the gas fees for the two transactions.
The second transaction involves various function calls to other contracts and will cost c.~15 USD equivalent of gas.
However, by the end of the attack, the attacker has c.~55k ETH, 50k MKR, and 145m DAI.
This amounts to a net profit of 191m USD.
Moreover, the attacker can design the attack smart contract such that the transaction is reverted if it becomes unprofitable. 
This makes the attack \emph{risk free} from a cost perspective for the attacker.
As pointed out above, the attacker can further create unlimited DAI to buy up existing liquidity on decentralized and centralized exchanges.

\section{DeFi Lending Protocols}
\label{sec:defilending}

After having presented a specific attack vector, we now turn to a generalization of the financial risk that exists for DeFi lending protocols.
This section provides a formal system model for a DeFi lending system and characterizes system constraints.
Appendix~\ref{appendix:defi-params} Table~\ref{tab:modelledprotocols} details the parameters for existing DeFi lending protocols that we seek to generalize in this section. 

\subsection{DeFi Lending Protocol Model}
\label{sec:defisystemmodel}

Overcollateralized borrowing allows an agent to provide an asset A as collateral to receive or create another asset B, of lower value, in return.
The asset B, typically together with the payment of a fee, can be returned and the agent redeems its collateral in return.
However, borrowed asset B may have different properties to asset A: for example, an agent might provide a highly volatile asset A and receive a price-stable asset B in return.
Furthermore, a third asset C can serve as a governance mechanism, such as MKR.
Holders of asset C are able to influence the rules of the DeFi lending protocol.
In absence of a governance asset, DeFi lending protocols typically replace this function with a central privileged operator introducing counter-party risk.

At the agent level, a DeFi lending protocol permits agents $k \in K$ to escrow units of cryptocurrency $i$, $c_i$ and borrow (or issue) units of another cryptocurrency $d$ against that value.
We formulate the constraints herein such that $i \in I$, where $I$ denotes all the permissible collateral types. 
Appendix~\ref{appendix:defi-params} Table~\ref{tab:modelledprotocols} provides the collateral assets and liquidation ratios for DeFi protocols that account for \definitionrelevance of the DeFi lending market. 
The prices of escrowed and borrowed assets are typically quoted with reference to an agreed quote currency, e.g. USD.


At the system level, a DeFi protocol is the aggregation of the individual acts of borrowing by agents, such that the system collateral of type $i$ is given by $C_i=\sum_{k=1}^{K}c_{i,k}$ for $K$ agents.
We formally define an economically secure DeFi lending protocol as follows:

\begin{definition}[Economically Secure DeFi lending protocol]
Assuming rational agents, a DeFi lending protocol is economically secure if it ensures that $\forall t$, with reference to a basis of value (e.g. USD), the total value of the system debt $D$ at time $t$ is smaller than the total value of all backing collateral types $I$ ($\sum_{i=1}^{I} C_i$) at time $t$.
\end{definition}

%

\subsection{Economic Security Constraints}

We now provide three constraints on the economic security of a DeFi lending protocol.
These constraints apply to DeFi protocols which feature one or several collateral assets and which may additionally feature a reserve asset.

\point{The Overcollateralization Constraint}\label{sec:oc-constraint}
Since the values of both the collateral assets and debt are subject to price fluctuations, overcollateralization seeks to ensure that there is \textit{always} sufficient collateral to cover debt, i.e., to avoid insolvency. 

\begin{definition}[Overcollateralization]
When escrowed collateral $c_i$ has a greater value with respect to a basis of value than the issued loan $d$.
\end{definition}

Denoting the overcollateralization factor as $\lambda_i \geq 0$, such that each collateral type has its own minimum collateralization ratio, and the price and quantity of an asset as $P()$ and $Q()$ respectively, the margin $M$ of overcollateralization at time $t$ at the system level\footnote{The system level perspective looks at the aggregates of assets and liabilities; depending on the protocol the ability to use one asset to cross-subsidize an undercollateralized other asset may be restricted.} is as follows (summing over agents $k \in K$ and collateral types $i \in I$). A protocol designer faces a trade-off. If the parameter $\lambda_i$ is too low, volatile markets may mean that the protocol becomes undercollateralized. 
However, if it is too high, then there is significant capital market inefficiency, with more capital than necessary in escrow, leading to opportunity costs of capital.
\begin{equation}
\label{eqn:systemmargin-vanilla}
    M_t = (1+\lambda_i)\sum_{k=1}^K\sum_{i=1}^I P_{c_{i,k,t}}Q_{c_{i,k,t}}-\sum_{k=1}^K d_{k,t}
\end{equation}

Clearly, $M_t \geq 0 \iff \sum_{k=1}^K d_{k,t} \leq (1+\lambda_i)\sum_{k=1}^K\sum_{i=1}^I P_{c_{i,k,t}}Q_{c_{i,k,t}}$.
Should $M<0$, then the margin of overcollateralization is negative and therefore the system as a whole is undercollateralized.

In addition, a protocol may have another pool of reserve liquidity available, enabling it to act as a lender of last resort.\footnote{If a protocol does not have this, $Q_{\Pi,t}=0$.}
For example, one such pool of collateral could be constituted by governance tokens $\Pi$ for the protocol itself.\footnote{MKR tokens in the case of Maker.}
In a DeFi protocol, participants can have voting power in proportion to the number of governance tokens they hold. 
The total value of this pool of collateral is given by $P(\Pi)Q(\Pi)$, and thus adding this into the margin of overcollateralization for the system yields:
\begin{equation} 
\label{eqn:systemmargin-reserve}
M_t = (1+\lambda_i)\sum_{k=1}^K\sum_{i=1}^I P_{c_{i,k,t}}Q_{c_{i,k,t}} + P_{\Pi,t}Q_{\Pi,t} - \sum_{k=1}^K d_{k,t} 
\end{equation}

Therefore, at the system level, the necessary condition for economic security in terms of overcollateralization is $M_t \geq 0$.
In the event that $(1+\lambda_i)\sum_{k=1}^K\sum_{i=1}^I P_{c_{i,k,t}}Q_{c_{i,k,t}} < \sum_{k=1}^K d_{k,t}$, the reserve asset $\Pi$ of a protocol can be used as a \textit{lender of last resort} to buy the collateral value.
If $M < 0$ even the liquidation of all of the primary collateral asset and reserve asset would be insufficient to cover the total system debt.
Since tt is possible that the collateral and reserve assets are correlated\footnote{There is evidence that crypto-assets display high intra-class correlation, limiting the advantage of diversification~\cite{koutsouri2019}.}, the ability of a reserve asset to recapitalize a system may be limited in the event of a sharp price drops. 
Absent any additional protocol specific defense mechanisms, this would constitute a catastrophic system failure since the borrowed funds would become worthless as they would no longer be redeemable.

\point{The Liquidity Constraint} \label{sec:liq-constraint}
In an illiquid market, liquidating a collateral asset may only be possible with a significant \textit{haircut}, where the collateral is sold at a discount.
Following~\cite{nikolaou2009liquidity}, we define market liquidity as follows. 

\begin{definition}[Market liquidity]
    A measure of the extent to which a market can facilitate the trade of an asset at short notice, low cost and with little impact on its price. 
\end{definition}

The liquidity available in a market implies a security constraint: in expectations, over a certain time horizon, DeFi marketplaces can offer enough liquidity that in the event of a sustained period of negative price shocks, a protocol will be able to liquidate its collateral quickly enough to cover its outstanding debt liabilities.

For a time interval $[0,T]$ this can be expressed as:

\begin{equation} \label{eq:liquidityconstraint}
\int_{0}^{T} \mathbb{E}[\Omega] d \Omega \leq \mathbb{E}[\Omega_{max}]
\end{equation}
where $\Omega$ denotes the total notional traded value, i.e., the (average) price multiplied by the quantity for each trade.
For a given trade $\omega$ of size $q$, $\omega = \bar{p}q$; aggregating these trades for a total number of trades $J$ provides $\Omega = \sum^{J}_{j \in J}\omega_j$.
$\Omega_{max}$ denotes the maximum notional value that could be sold off during a period of distress in the financial markets.

In the event of a severe price crash, on the assumption that a protocol is collateralized to a representative~$150\%$, we assume a protocol will seize~100\% of the debt value from the collateral pool, and seek to sell this collateral as quickly as possible on a market pair to the debt asset.
Once a buyer has traded the debt asset $d$ for the collateral, the protocol could then burn the debt $d$, effectively taking it out of circulation, offsetting the liability.
Therefore, the impact that negative price shocks would have on a DeFi lending protocol, and how quickly they materialize, depend on liquidity available on all collateral/debt pairs.
In the event of a liquidity crisis, the demand for liquidity outstrips supply\footnote{Indeed, such liquidity crises were at the heart of the Financial crisis of 2007-8, as the value of many financial instruments traded by banks fell sharply without buyers~\cite{guardian_2012}.}, such that equation (\ref{eq:liquidityconstraint}) is binding.
Indeed, if equation \ref{eq:liquidityconstraint} is binding there are not enough buyers in the market to buy the ETH that is for sale.

\point{The Counterparty Risk Constraint}
DeFi lending protocols are not fully decentralized on account of, for instance, the possibility of oracle attacks (which could cause a flash-crash), as well as privileged access to the smart contracts. 
Therefore it is necessary to either assume the ``operator'' of the protocol is honest, or that the operator only offers the services of the protocol provided they are profitable for them.
We formally model this counterparty risk by assuming that its existence in a given protocol creates a risk premium, $\psi$, such that for an agent deciding between earning a return in a DeFi lending protocol vs elsewhere, the expected return in the DeFi protocol ($r_D$), once adjusted for the risk premium ($\psi$), must be higher than an outside return $r_f$. 
Formally, we have participation constraint $r_D-\psi > r_f$.
This constraint is a participation constraint, and in Section~\ref{sec:stress-test} we assume that this inequality holds, such that agents have already chosen to participate in the protocol.

There exists an inherent trade-off in counterparty risk.
On the one hand, governance mechanisms implemented through voting allow for a certain degree of decentralization whereby multiple protocol participants can influence the future direction of a protocol.
Depending on the distribution of tokens, this may reduce the risk of one party becoming malicious.
However, it also opens the door to attacks on the voting system, as we introduced in Section~\ref{sec:governance-attack}.
On the other hand, a single `benevolent dictator' who controls the governance mechanism can prevent the attacks introduced in Section~\ref{sec:governance-attack}.
Yet this requires trusting that this central entity does not lose or expose its private keys controlling access to the smart contracts governing the protocol and that this central party cannot be bribed to behave maliciously.

\section{Stress-Testing DeFi Lending}
\label{sec:stress-test}

This section considers the financial security of a generic DeFi lending protocol, stress-testing the architecture to quantitatively assess its robustness as inspired by central banks~\cite{bankofenglandstresstest,fedstresstest}.

\subsection{Stress-Testing Framework}

Central banks conduct stress tests of banking systems to test their ability to withstand shocks. 
For example, in an annual stress test, the Bank of England examines what the potential impact would be of an adverse scenario on the banking system~\cite{bankofenglandstresstest}.
The hypothetical scenario is a ``tail-risk'' scenario, which seeks to be broad and severe enough to capture a range of adverse shocks. 
Following such best practice, we devise and implement a stress-test of the DeFi architecture.

\subsection{Simulation Approach}\label{sec:assumptions}

We leverage the generic DeFi lending protocol architecture as developed in Section~\ref{sec:defisystemmodel}.
We focus on a single collateral asset here for tractability, but this analysis can be extended lending protocols which rely on overcollateralization by multiple volatile collateral assets in combination with reserve assets.
In part reflecting Appendix~\ref{appendix:defi-params} Table~\ref{tab:modelledprotocols}, we make the following assumptions about the initial state of the system.
\begin{enumerate}
    \item The lending protocol allows users to deposit ETH as their single source of collateral $c_i$.
    \item The lending protocol has 1m tokens of a generic reserve asset, which at the start of the simulation has the same price as ETH but with exactly half of the historical standard deviation of ETH taken over the sample period.
    \item By arbitrage among borrowers, before the crash the lending protocol as a whole is collateralized to $\lambda_i + \epsilon$, i.e., just above the minimum collateralization ratio.
    \item At the start of the crisis, the protocol has a collateralization ratio of exactly~150\%, such that every USD of debt is backed by~1.50 USD of collateral.
    \item Each unit of debt $d^k$ maintains a peg of~1:1 to the US dollar, allowing us to abstract from the dynamics of maintaining the peg.
    \item At the start of the sell-off, it is possible to sell~30,000 ETH per day without having an impact on price.\footnote{This assumption is based on the~24-hour volume of ETH/DAI across markets listed on CoinGecko on February~7th,~2020, and as such is only a rough proxy for the market liquidity. We use this figure only as a baseline for parameterization and to highlight the theoretical possibility of illiquidity causing default.}.
    \item The amount of reserve asset $\Pi$ is fixed at the start of the sell-off at 1m units.
    \item System debt levels range from~100m USD to~400m USD, seeking to approximately reflect the levels of capital escrowed in DeFi protocols as in Appendix~\ref{appendix:defi-params} Table~\ref{tab:defiprojects}.
\end{enumerate}
Next, we detail the methodology we follow to obtain our simulation results. 

\point{Price simulation}
Firstly, we obtain OHLCV data at daily frequency~\cite{cryptocompare}, focusing on the period January~1st,~2018 to February~7th,~2020, incorporating the large fall in the ETH price in early~2018.
We present the evolution of ETH close prices in Appendix~\ref{sec:price-data} Fig.~\ref{fig:ethprices} and a histogram of log returns in Appendix~\ref{sec:price-data} Fig.~\ref{fig:ethreturns}.
Perhaps the most notable element is the decline in the ETH/USD price over the course of~2018, with the price of ETH falling from an all-time-high of 1,432.88 USD to c.~220 USD as of February~7th,~2020.
Taking parameters from this historical data\footnote{For the daily ETH/USD price data we find mean log returns of $0.001592$ and standard deviation $0.050581$, parameter values which have been independently verified.}, we use Monte Carlo simulation
to capture how the ETH and reserve prices may be expected to evolve over the next 100 days.
Monte Carlo simulation leverages randomness to produce a range of outcomes of a stochastic system.
We simulate~5,000 randomly generated paths, using a geometric Brownian motion, specified with the following equation.

\begin{equation}
    P_{c_{i,t}} = P_{c_{i,0}} \exp \left[(\mu_i - \frac{\sigma_i^2}{2})t + \sigma_i W_t \right]
\end{equation}

$W_t$ denotes a Wiener process~\cite{wiener1976collected} and for collateral type $i$ $\mu_i$ denotes the drift and $\sigma_i$ denotes the volatility.\footnote{In this estimation, we draw shocks from the normal distribution, as is standard in GBM. Since performing a Jarque-Bera test~\cite{jarque1987test} over the sample period suggests that the log-returns are non-normal, it is possible that in our estimation we underestimate the impact of heavy tails. Therefore, we present a best-case upper bound; in practice, undercollateralization could precipitate more quickly.}
Of the~5,000 simulations, our subsequent analysis is focused on the iteration which yields the fastest undercollateralization event. 
By focusing on this worst-case, we test the DeFi lending protocol with a ``black swan'' event, representing a severe challenge to its robustness.

\point{System simulation}
We propose a simple model for the decline in liquidity over time as follows.
\begin{equation}
    L = L_{0} \exp(- \rho t)
\end{equation}
where $L_0$ denotes the initial amount of ETH that can be sold per day.
Intuitively, this equation captures the notion that in the event that the protocol attempts to sell large volumes each period, the amount of liquidity available in the next period will be lower.

In this simulation approach, we make a simplification by not modeling the impact that selling large volumes of collateral will have on the price of the collateral asset.
It is highly likely that in such a sell-off scenario, the selling of large volumes would serve to endogenously push the price lower.
Therefore what we present here represents an upper bound on the price behavior: in reality, the price drop may be even worse than the one we examine. 

\subsection{Simulation Results}

We start with the Monte Carlo simulation of the correlated asset paths, before considering the impact this would have on a DeFi lending protocol and an ecosystem of multiple lending protocols.

\point{Monte Carlo Price Simulation}
To capture the effects of different correlations between the collateral asset and the reserve asset, we consider three different extents of correlation between the collateral and reserve asset: 
(i) strong, positive correlation~(0.9), 
(ii) weak, positive correlation~(0.1) and 
(iii) strong negative correlation~(-0.9).
We then generate correlated asset paths during the Monte Carlo simulation process. 
In this section we report results for strong correlation, but include those for weak correlation and strong negative correlation in Appendix~\ref{appendix:sim-results}.

Fig.~\ref{fig:ethmontecarlo} shows the results of~5,000 runs of the Monte Carlo simulator for the ETH price, and Appendix~\ref{appendix:sim-results} Fig.~\ref{fig:resmontecarlo} shows the results for the reserve asset price in the presence of strong positive correlation in the asset price returns. 
The starting prices of assets as used in the simulator is the close price of ETH/USD on February~7th,~2020.

\begin{figure}[tb]
    \centering
    \includegraphics[width=\columnwidth]{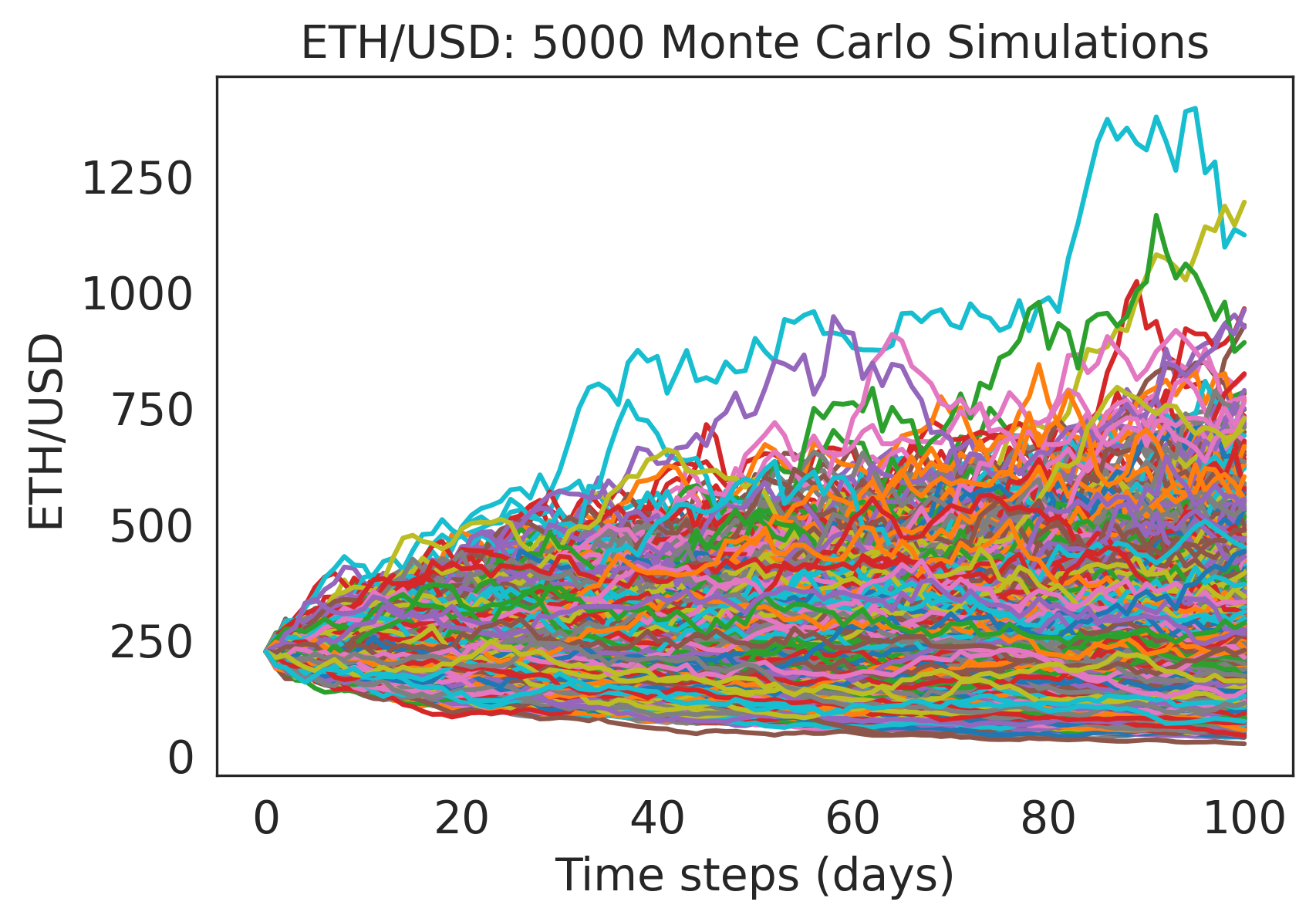}
    \caption{Monte Carlo forecast of ETH prices over the next 100 days from February~7th,~2020.}
    \label{fig:ethmontecarlo}
\end{figure}

We isolate the simulation which yields the fastest undercollateralization event.\footnote{We plot the co-evolution of the asset price paths for strong correlation in Appendix~\ref{appendix:sim-results} Fig.~\ref{fig:correlated-sims-strong}.}
In Appendix~\ref{appendix:sim-results} Fig.~\ref{fig:correlated-sims-strong} it is clear that in this worst case scenario for the ETH/USD price, the price of the reserve asset similarly falls.
This illustrates the risk of using a reserve asset which is positively correlated with the collateral asset: if the price of the collateral asset falls, relative to the same basis of value the reserve asset value is likely to fall, limiting the ability of a DeFi lending protocol to recapitalize itself. 

\begin{figure*}[ht]
    \centering
    \includegraphics[width=\textwidth]{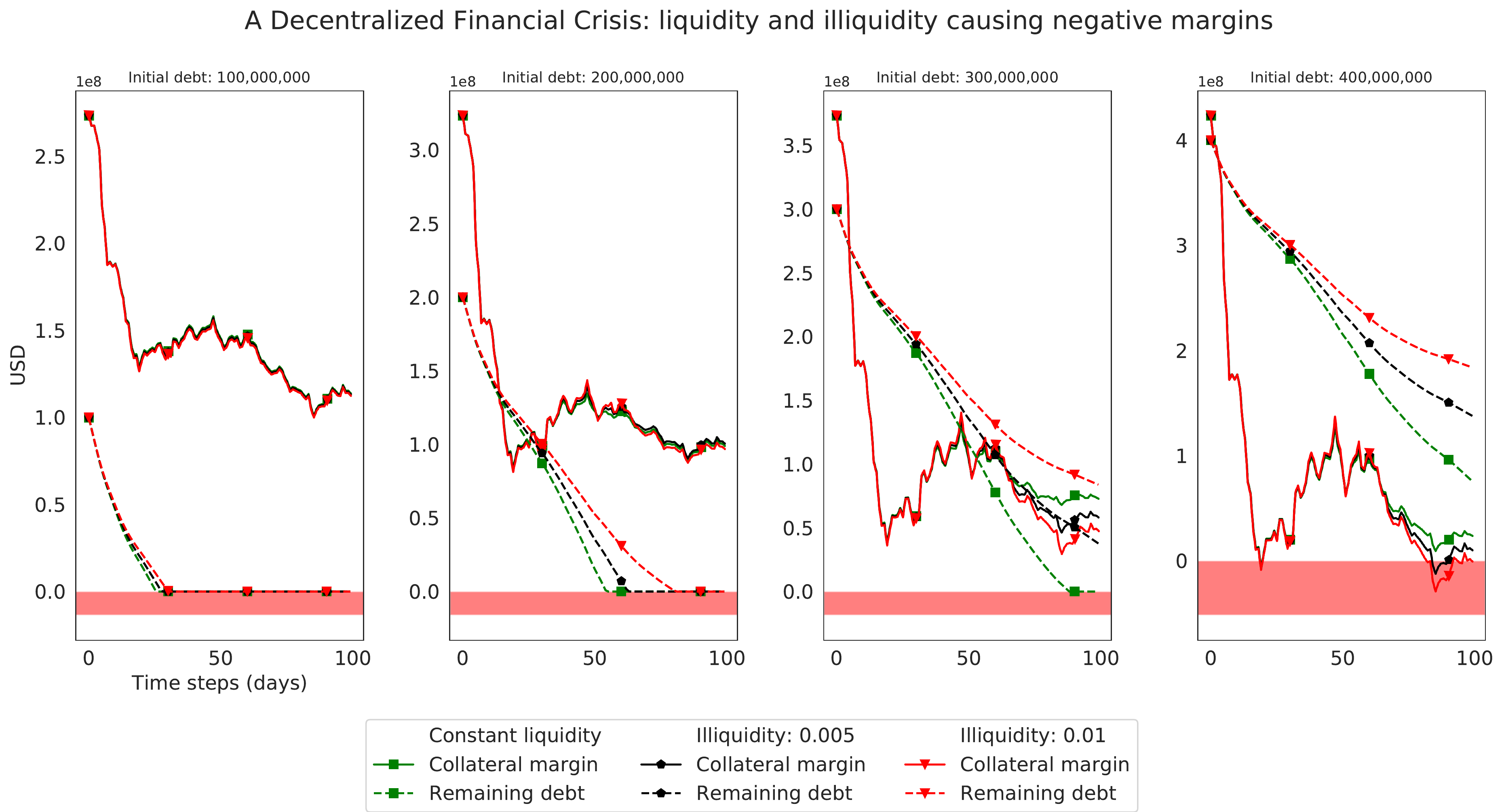}
    \caption{A DeFi lending protocol experiencing a sharp decline in the price of its collateral and reserve assets. Panels correspond to four different levels of system debt, with each panel showing the evolution of the collateral margin (solid lines) and the total debt outstanding (dashed lines). Each panel also shows the consequences of different liquidity parameters. The margin becomes negative in panels 3 and 4--- entering the red region below zero---the situation in which the lending protocol has become undercollateralized.}
    \label{fig:final-simulation-strong-positive}
\end{figure*}

\point{Impact on Collateral Margin}
We take the simulation yielding the fastest undercollateralization event and consider the impact this would have on the collateral margin of a DeFi lending protocol. 
The main results of this are presented in Fig.~\ref{fig:final-simulation-strong-positive}.
Plotted with solid lines is the evolution of the total collateral margin (comprising the collateral and the reserve asset) over time as the prices of the collateral asset and reserve asset decline. 
The dashed lines indicate how the amount of system debt evolves through time, on the assumption that at the start of the 100 day period, the protocol seeks to sell off all of the debt.
The speed at which the debt can be liquidated through the sale of its backing collateral in turn depends on the available liquidity in the market for which we consider~3 cases: 
\begin{enumerate}
    \item constant liquidity (such that it is possible to sell a constant amount of ETH every day at the average daily price)
    \item  mild illiquidity (where the illiquidity parameter is arbitrarily set to some low level $\rho=0.005$)
    \item illiquidity, such that $\rho=0.01$.
\end{enumerate}

Where the initial system debt level is 100m USD, regardless of the liquidity parameter, the collateral margin does not become negative.
However, at higher levels of debt, we see that the margin gets closer to~0, and once the debt level reaches~400m USD, the margin does indeed fall below 0, such that the protocol is undercollateralized overall.
In the fourth panel of Fig.~\ref{fig:final-simulation-strong-positive} we see that after 19 days of the protocol attempting to liquidate as much debt as possible, due to illiquidity it is unable to liquidate in time and the margin becomes negative. 
This would constitute a crisis in a DeFi protocol: each unit of debt would not have sufficient collateral backing, and rational agents would walk away from the protocol without repaying their debt.\footnote{In the event that strong-identities (i.e., where the mapping between an agent an an online identity is one-to-one and time invariant) are enforced on-chain, this calculus may change for agents, reducing the probability of a crisis by increasing the costs to the agent of reneging on their debt commitments. In this paper we proceed under the assumption that strong-identities are not enforced.}
Notably, the results presented in the Appendix~\ref{appendix:sim-results} show that a weakly correlated reserve asset is able to slow or prevent the collateral margin from becoming negative (see Appendix~\ref{appendix:sim-results} Fig.~\ref{fig:final-simulation-weak-positive}) while a strongly negative correlation between the assets is actually able to bolster the collateral margin (see Appendix~\ref{appendix:sim-results} Fig.~\ref{fig:final-simulation-strong-negative}).

\begin{figure}
    \centering
    \includegraphics[width=\columnwidth]{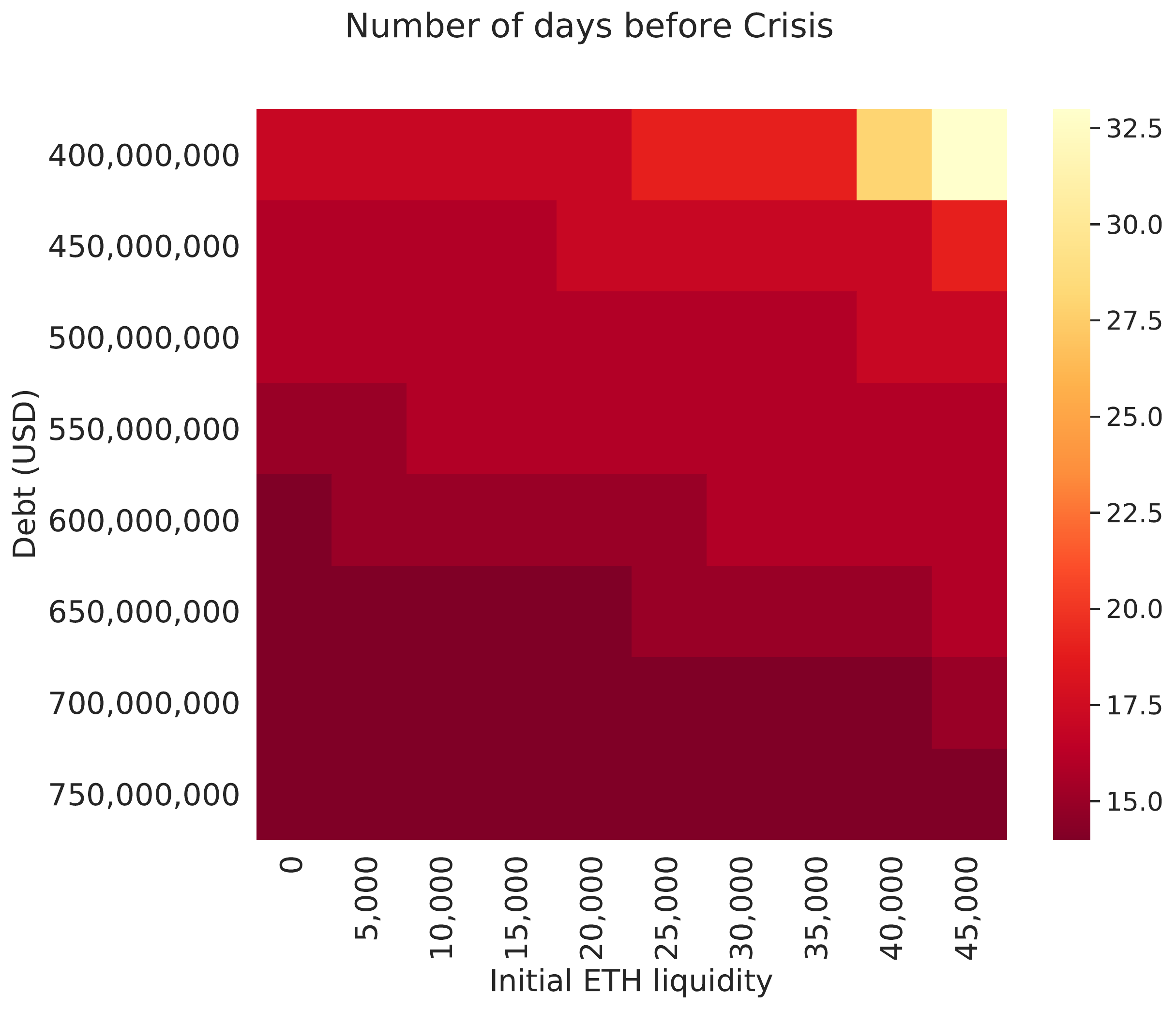}
    \caption{Number of days before the collateral margin becomes negative, depending on the amount of system debt and the initial amount of ETH that can be sold within 24 hours.}
    \label{fig:first-negative_vols}
\end{figure}

For the case where the collateral and reserve assets are strongly positively correlated, we consider how \textit{quickly} a crisis may materialize for varying starting values of ETH liquidity and initial debt in Fig.~\ref{fig:first-negative_vols}.
Fig.~\ref{fig:first-negative_vols} shows that for a given amount of debt, the lower the starting liquidity (i.e., the amount that can be sold within 24 hours), the faster a negative margin precipitates.
Similarly, for a fixed initial starting liquidity, the more debt there is in the system the faster the margin will become negative, down to below~15 days.

\section{Related Work}\label{sec:related-work}

There is a paucity of directly related work.
However, existing work can be divided into the following categories.
A series of fundamental results in relation to the ability of non-custodial stablecoins to maintain their peg is provided in \cite{klages2019stability}.
It is shown that stablecoins face deleveraging spirals which cause illiquidity during crises, and that stablecoins have \textit{stable} and \textit{unstable} domains.
The model primarily involves the assumption of two types of agents in the marketplace: the stablecoin holder (who wants stability), and the speculator (who seeks leverage). 
The authors further demonstrate that such systems are susceptible to tail volatility.
While unpublished, \cite{cao2018} uses option pricing theory to design dual-class structures that offer fixed income stable coins that are pegged to fiat currency.
Further, \cite{pentland2018digital} considers how one might build an asset-backed cryptocurrency through the use of hedging techniques.



\section{Conclusions}
\label{sec:conclusion}

This paper has sought to demonstrate that, as they stand, DeFi lending protocols are liable to a variety of attack vectors.
Firstly, we show the feasibility of an attack on the governance mechanism of Maker, finding that, prior to the fix implemented by Maker, provided an attacker was able to lock~27.5m USD of governance tokens they would have been able to steal all~0.5bn USD worth of collateral within two blocks.
Therein we presented a novel strategy that would have enabled an attacker to steal the collateral within two transactions without the need to escrow any assets.

Secondly, after providing formal constraints on the robust operation of a DeFi lending protocol, we use simulations to show that a for given parameters a DeFi lending protocol may become under-collateralized.
We describe the interrelation of market liquidity and outstanding debt, showing how the larger the debt, or the less liquid a market, the faster insolvency can occur. 
We also consider different levels of correlation between the collateral and the reserve asset in a DeFi lending protocol and show that having a reserve asset that is weakly positively correlated or indeed negatively correlated can help to ensure protocol solvency.

These two types of failure mode in a DeFi protocol are potentially mutually reinforcing.
If the collateral and reserve assets of a DeFi lending protocol experience a sharp decline in price, the cost of acquiring enough governance tokens to undertake the governance attack would also likely fall. 
Conversely, should an actor undertake a governance attack, this would plausibly send shock waves throughout the DeFi ecosystem, serving to reduce the price of the collateral asset, in turn making under-collateralization more likely. 


\vspace{-1mm}
\section*{Acknowledgments}
\vspace{-1mm}
The authors thank the reviewers for their valuable
comments to improve the paper, and MakerDAO for their feedback.
This work has been partially supported by EPSRC Standard Research Studentship (DTP) (EP/R513052/1) and the Tezos Foundation.
\bibliographystyle{plain}
\bibliography{references}

\appendix
\section{Appendix}
\label{sec:appendix}

\subsection{Existing DeFi protocols}\label{appendix:existing-defi-protocols}

\begin{table}[H]
\setlength{\tabcolsep}{5pt}
\centering
\scriptsize
\begin{tabular}{llrl}
\toprule
 & \bf Project & \bf Capital~(USD) & \bf Blockchain \\\midrule
 \textbf{Lending} & Maker~\cite{makerdao} & \makercapitalization & Ethereum \\
 & Compound~\cite{compoundfinance} & \compoundcapitalization & Ethereum \\
 & Aave~\cite{aave} & \aavecapitalization & Ethereum \\
 & & & \\ \midrule
 \textbf{DEX} & Uniswap~\cite{uniswap} & \uniswapcapitalization & Ethereum \\
 & Bancor~\cite{bancor} & \bancorcapitalization & Ethereum \\
 & Kyber~\cite{kyber} & \kybercapitalization & Ethereum \\
  & & & \\ \midrule
 \textbf{Derivatives} & Synthetix~\cite{synthetix} & \synthetixcapitalization & Ethereum \\
 & Nexus~\cite{nexusmutual} & \nexuscapitalization & Ethereum \\
 & Erasure~\cite{erasure} & \erasurecapitalization & Ethereum \\
 & & & \\ \midrule
 \textbf{Payments}  & Lightning~\cite{lightning} & \lightningcapitalization & Bitcoin \\
 & Connext~\cite{connext} & \connextcapitalization & Ethereum \\

 & & & \\ \midrule
 \textbf{Assets}  & token Sets~\cite{tokensets} & \tokensetscapitalization & Ethereum \\
 & WBTC~\cite{wbtc} & \wbtccapitalization & Ethereum \\
 & Melon~\cite{melon} & \meloncapitalization & Ethereum \\ \bottomrule
\end{tabular}
\vspace{1mm}
\caption{Existing DeFi projects~\cite{defipulse} (\todaysdate{}).} \label{tab:defiprojects}
\end{table}

\subsection{Governance Attack on Maker}
\label{appendix:governance-attack-on-maker}

\begin{figure}[H]
\centering
\includegraphics[width=\columnwidth]{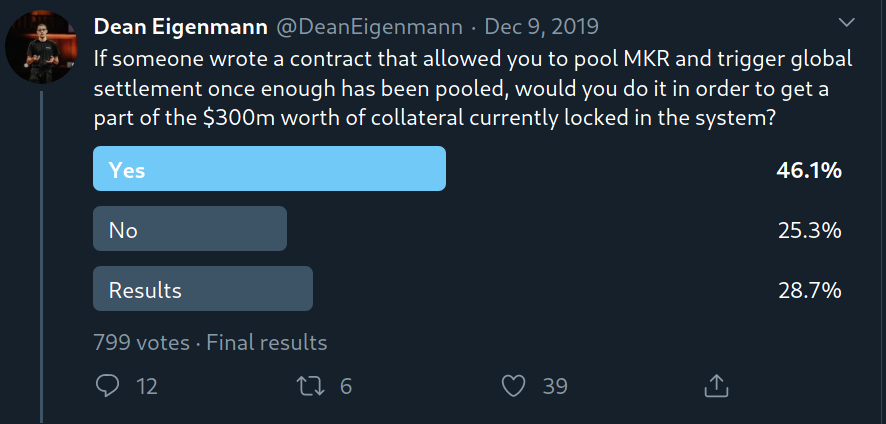}
\caption{Twitter poll for of a crowdfunding attack on MKR governance.}
\label{fig:mkr-crowdfunding-tweet}
\end{figure}

\begin{figure}[H]
\centering
\includegraphics[width=\columnwidth]{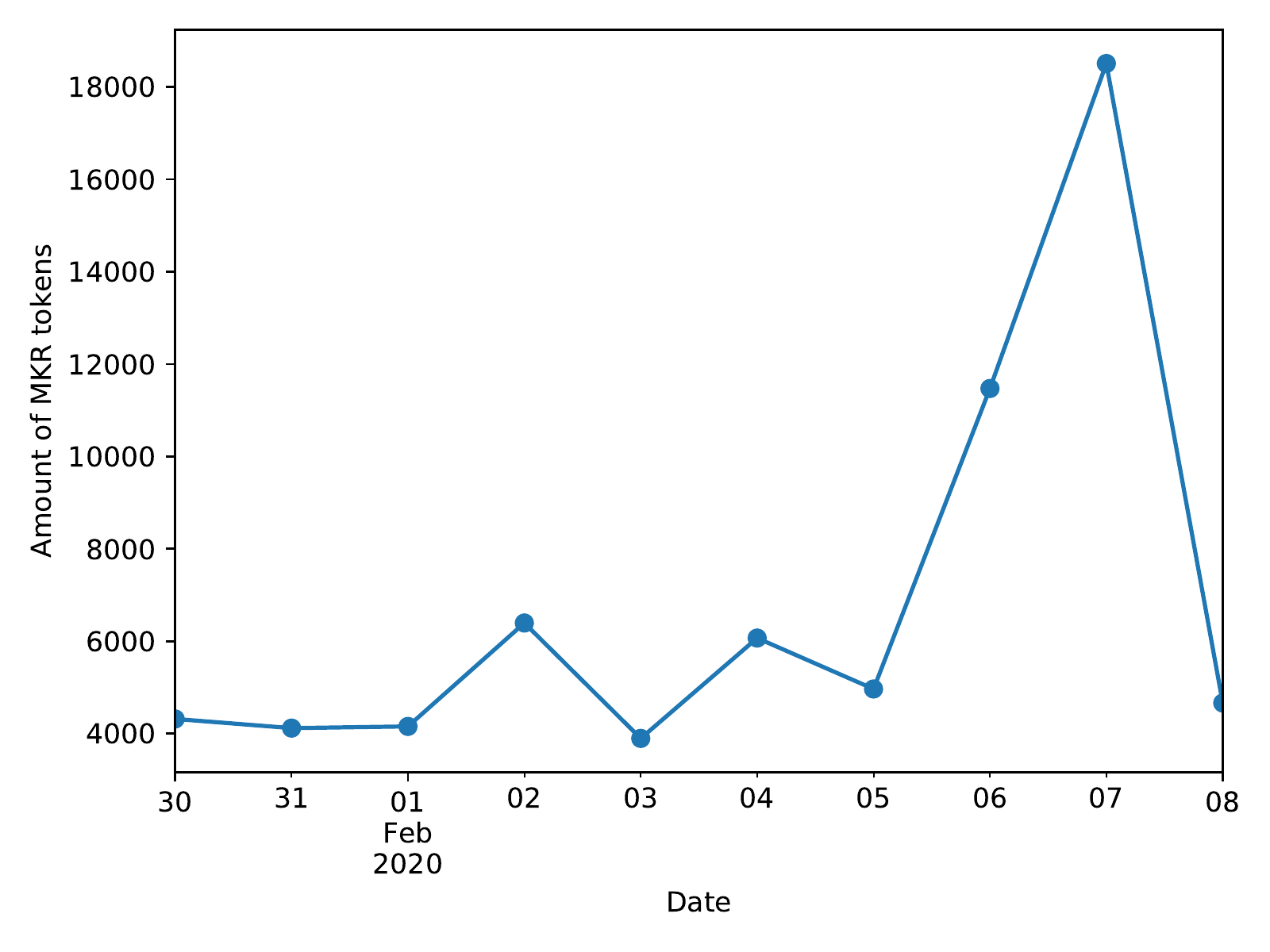}
\caption{Daily traded volume of MKR tokens between 2020-01-30 and 2020-02-08.}
\label{fig:mkr-traded-volume}
\end{figure}

\subsection{Parameters of DeFi lending platforms}
\label{appendix:defi-params}

\begin{table}[H]
    \setlength{\tabcolsep}{3pt}
    \centering
    \scriptsize
    \begin{tabularx}{\columnwidth}{lXl}
    \toprule
     \bf Protocol & \bf Collateral asset & \bf Reserve \\
     & \bf (liquidation ratio) & \bf asset \\
     \midrule
     Maker~\cite{makerdao} & ETH (150\%), BAT (150\%), USDC (125\%) & MKR \\
     Compound~\cite{compoundfinance} & ETH (133\%), BAT (167\%), DAI (133\%) & \\
     & REP (250\%), USDC (133\%), ZRX (167\%) & \\
     Aave~\cite{aave} & DAI (125\%), USDC (125\%), TUSD (125\%) &  \\
     & ETH (125\%), LEND (154\%), BAT (154\%) &  \\
     & KNC (154\%), LINK (143\%), MANA (154\%) &  \\ 
     & MKR (154\%), REP (154\%), WBTC (154\%) &  \\ 
     & ZRX (154\%) &  \\ 
     dYdX~\cite{dydx} & ETH (115\%), USDC (115\%), DAI (115\%)  &  \\
    \bottomrule
    \end{tabularx}
    \vspace{1mm}

    \caption{Parameters of DeFi lending platforms, comprising \definitionrelevance of DeFi market as of \todaysdate.} 
    \label{tab:modelledprotocols}
\end{table}

\subsection{Price data}
\label{sec:price-data}

\begin{figure}[H]
    \centering
    \includegraphics[width=\columnwidth]{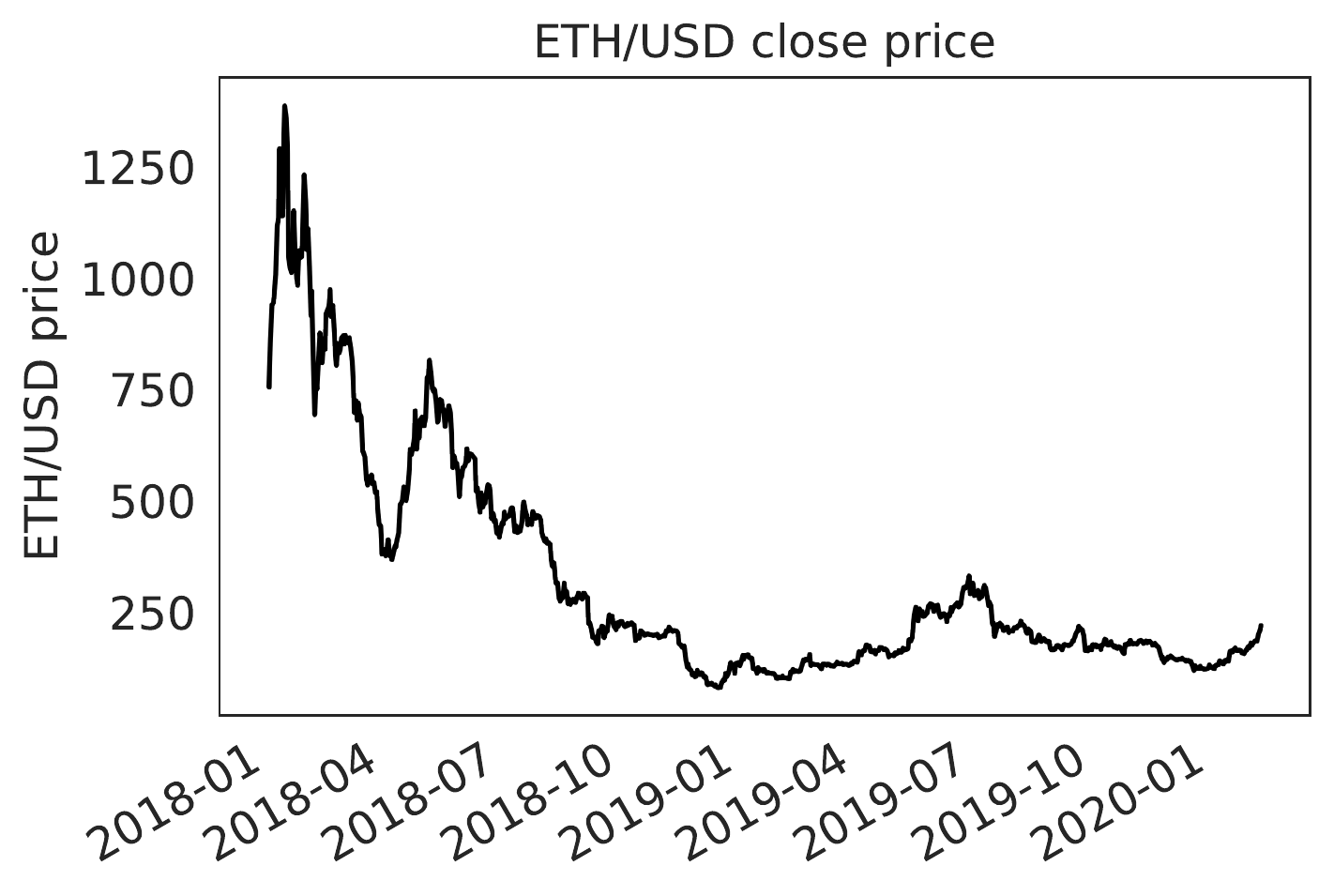}
    \caption{Close prices for ETH/USD over the period January~1st,~2018 to February~7th,~2020.}
    \label{fig:ethprices}
\end{figure}

\begin{figure}[H]
    \centering
    \includegraphics[width=\columnwidth]{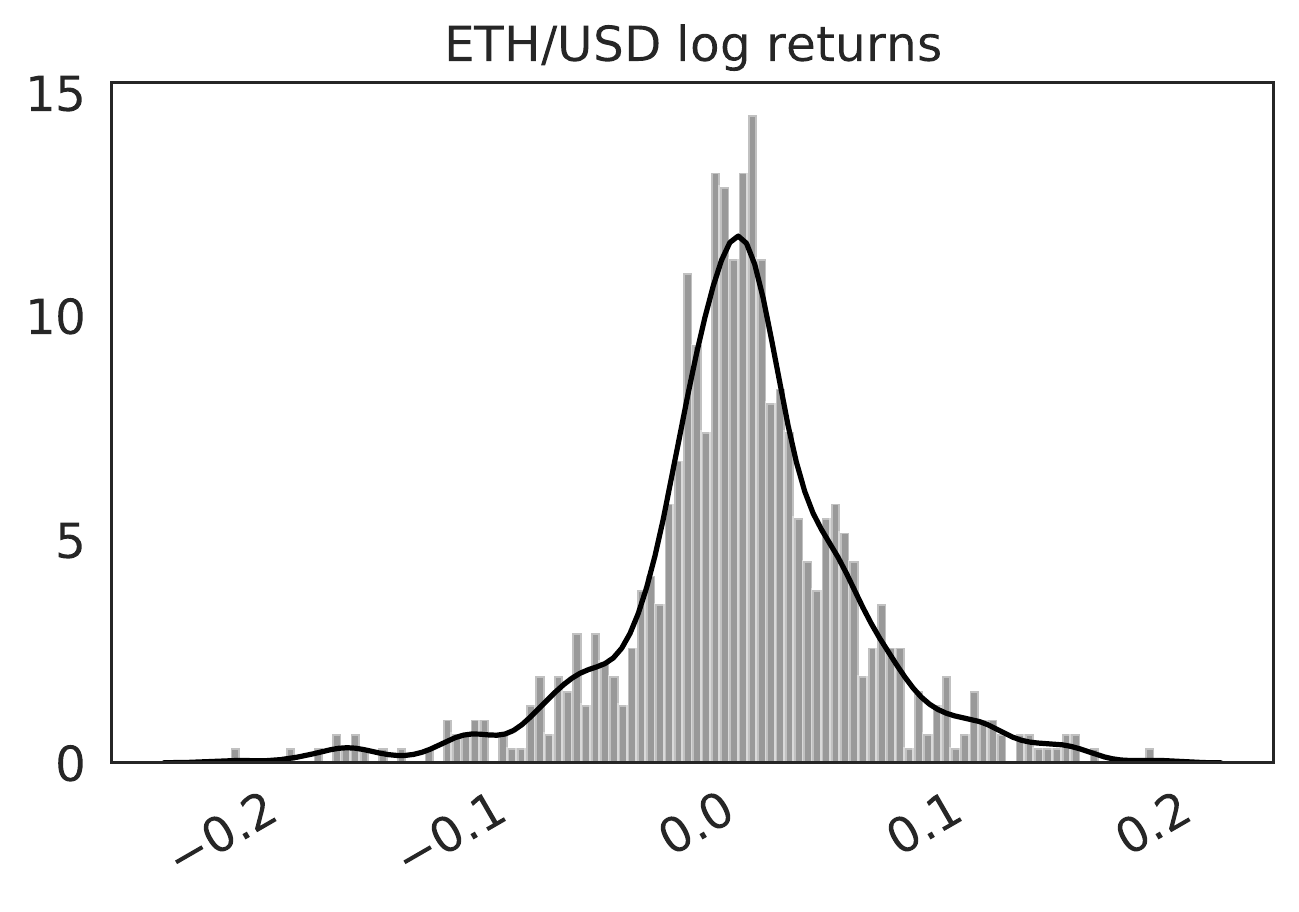}
    \caption{Log returns for ETH/USD over the period January~1st,~2018 to February~7th,~2020.}
    \label{fig:ethreturns}
\end{figure}

\newpage
\subsection{Simulation results}
\label{appendix:sim-results}

\begin{figure}[H]
    \centering
    \includegraphics[width=\columnwidth]{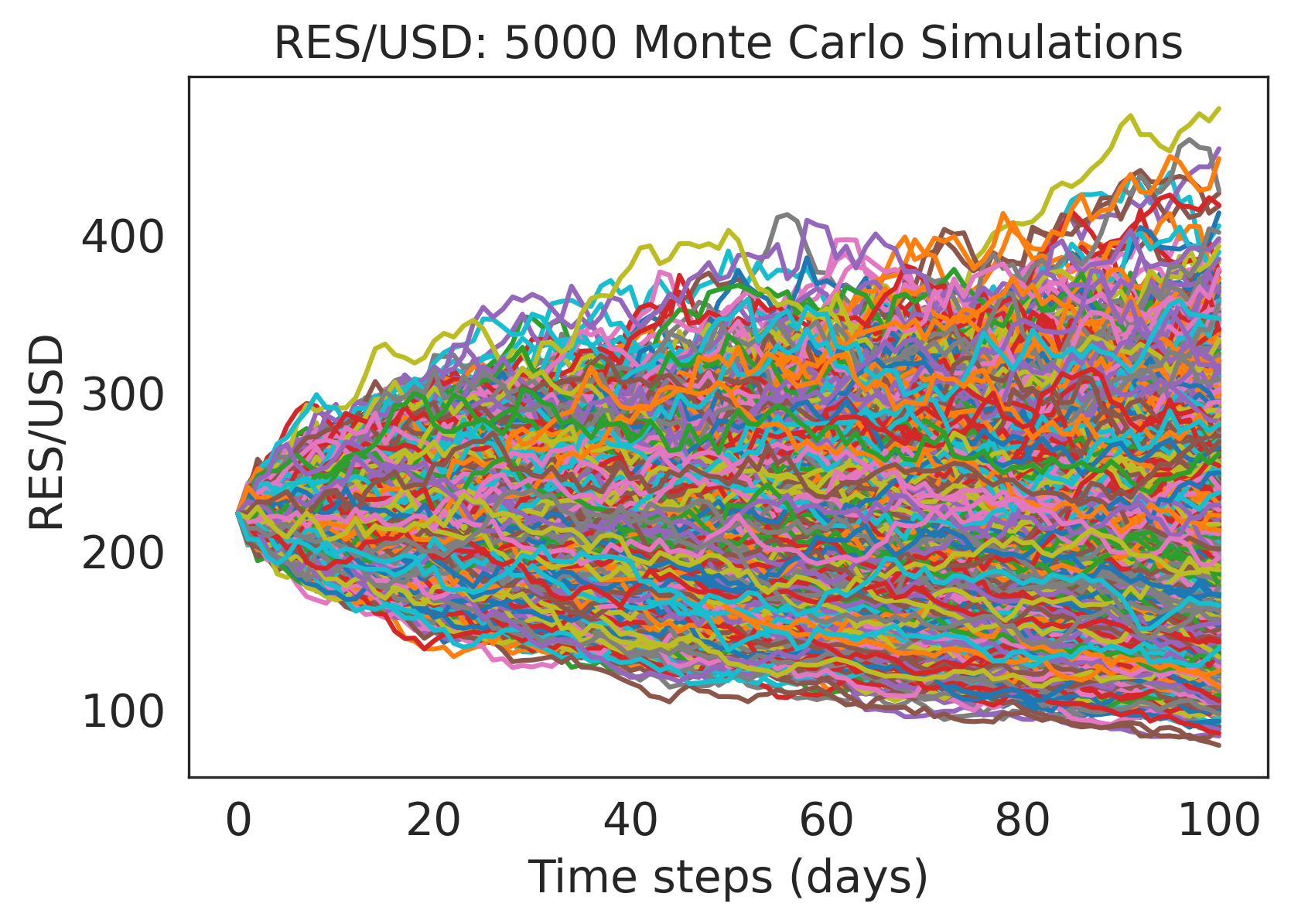}
    \caption{Monte Carlo forecast of the reserve asset price over the next~100 days from February~7th,~2020.}
    \label{fig:resmontecarlo}
\end{figure}

\begin{figure}[H]
    \centering
    \includegraphics[width=0.95\columnwidth]{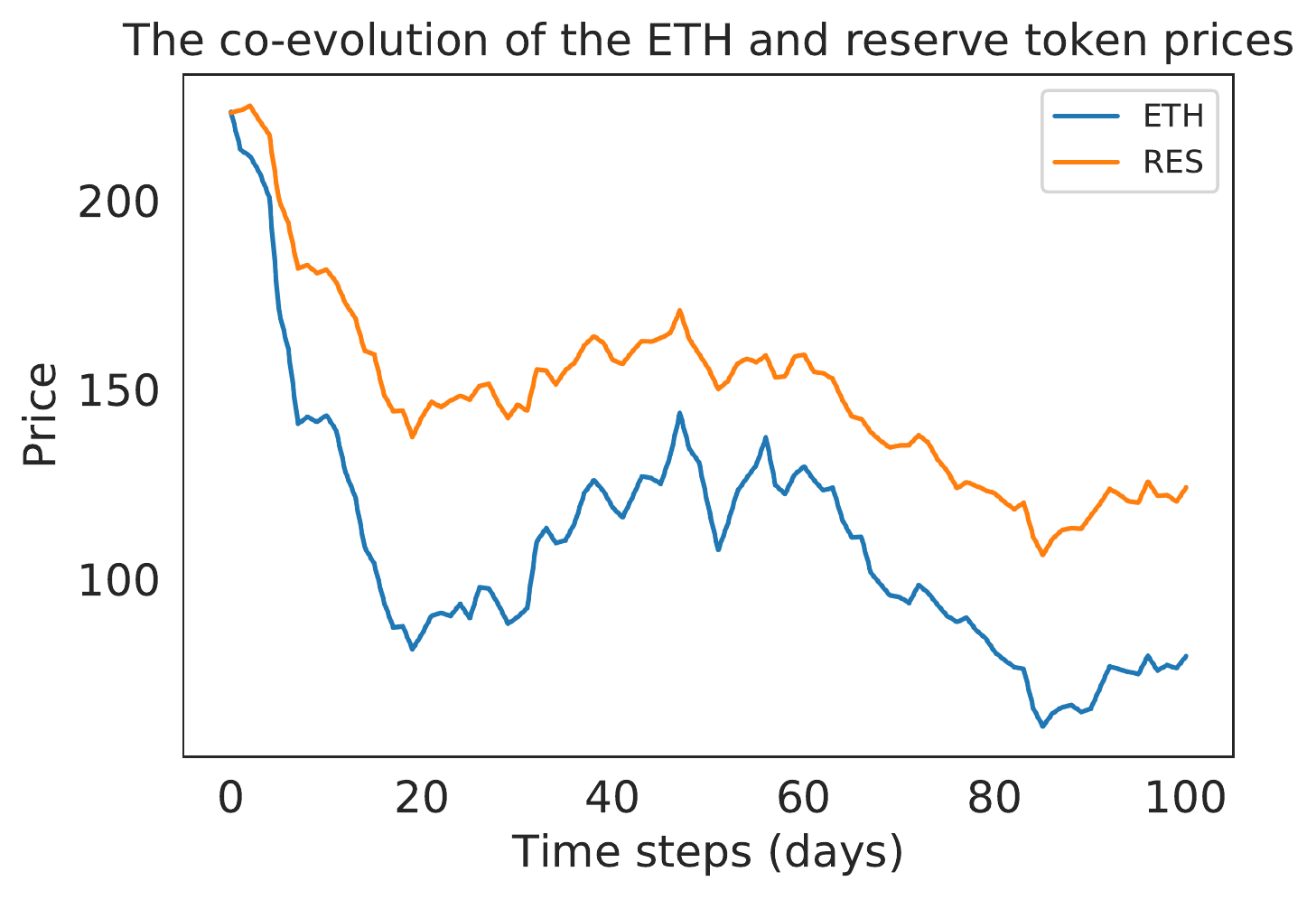}
    \caption{For the simulation yielding the fastest undercollateralization event, the co-evolution of the ETH and reserve asset prices where the asset price returns are strongly positively correlated.}
    \label{fig:correlated-sims-strong}
\end{figure}



\begin{figure*}[h]
    \centering
    \includegraphics[width=\columnwidth]{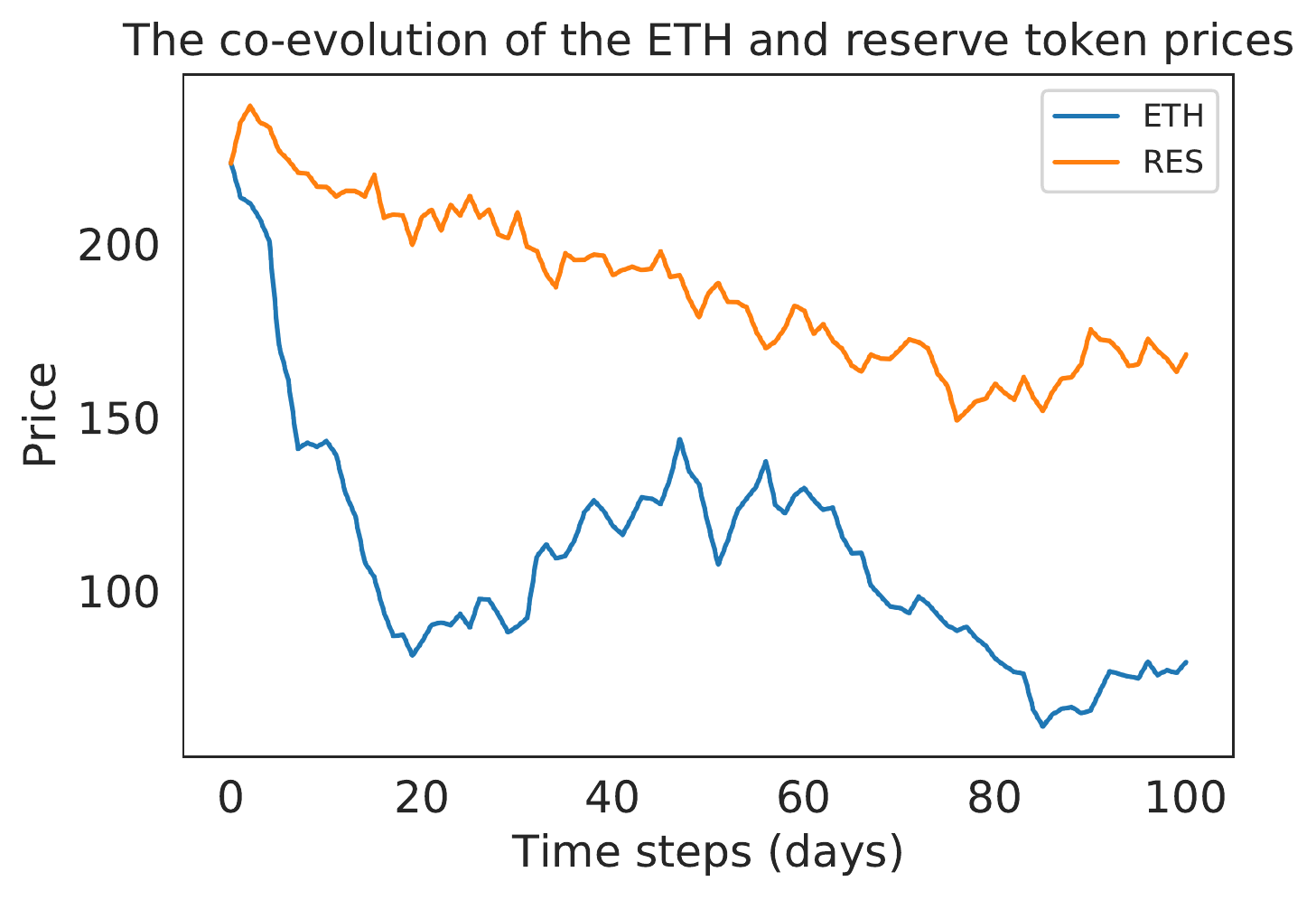}
    \caption{For the simulation yielding the fastest undercollateralization event, the co-evolution of the ETH and reserve asset prices where the asset price returns are weakly positively correlated.}
    \label{fig:correlated-sims-weak}
\end{figure*}


\begin{figure*}[h]
    \centering
    \includegraphics[width=0.95\textwidth]{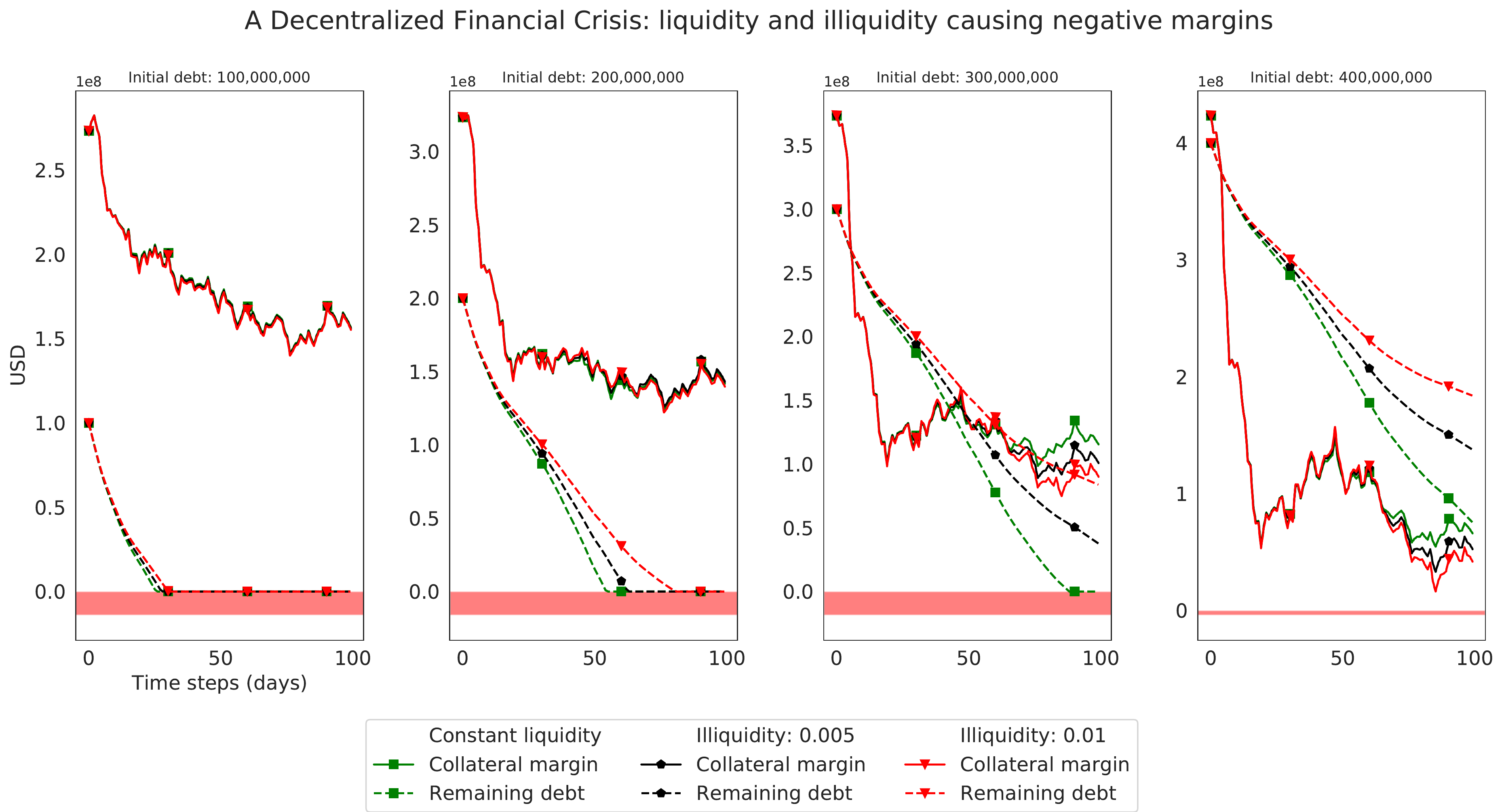}
    \caption{A DeFi lending protocol experiencing a sharp decline in the price of its collateral and reserve assets, where the assets have a correlation of 0.1. Panels correspond to 4 different levels of system debt, with each panel showing the evolution of the collateral margin and the total debt outstanding. Each panel also shows the consequences of different liquidity parameters.}
    \label{fig:final-simulation-weak-positive}
\end{figure*}



\begin{figure*}[h]
    \centering
    \includegraphics[width=\columnwidth]{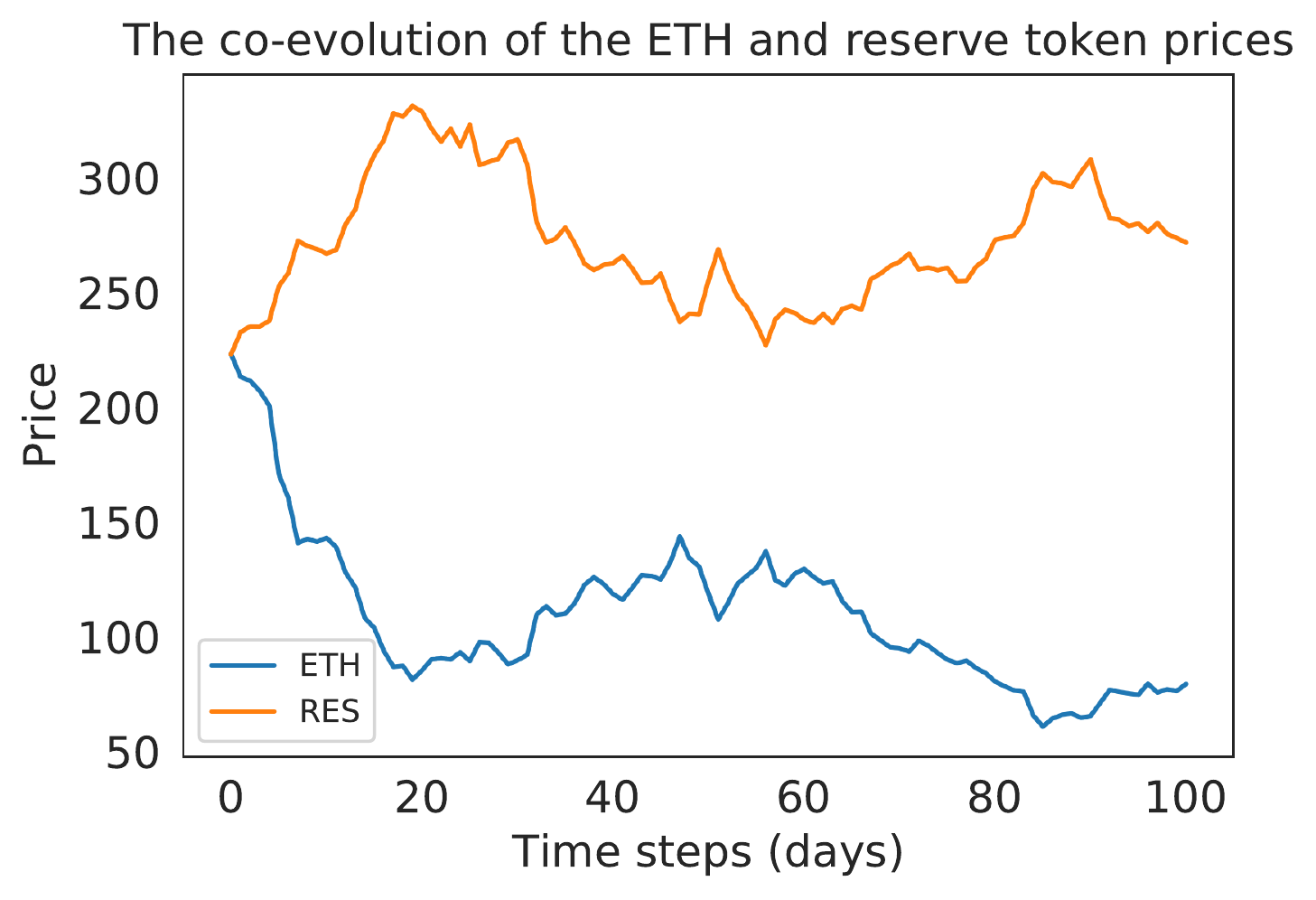}
    \caption{For the simulation yielding the fastest undercollateralization event, the co-evolution of the ETH and reserve asset prices where the asset price returns are strongly negatively correlated.}
    \label{fig:correlated-sims-strong-negative}
\end{figure*}


\begin{figure*}[h]
    \centering
    \includegraphics[width=0.95\textwidth]{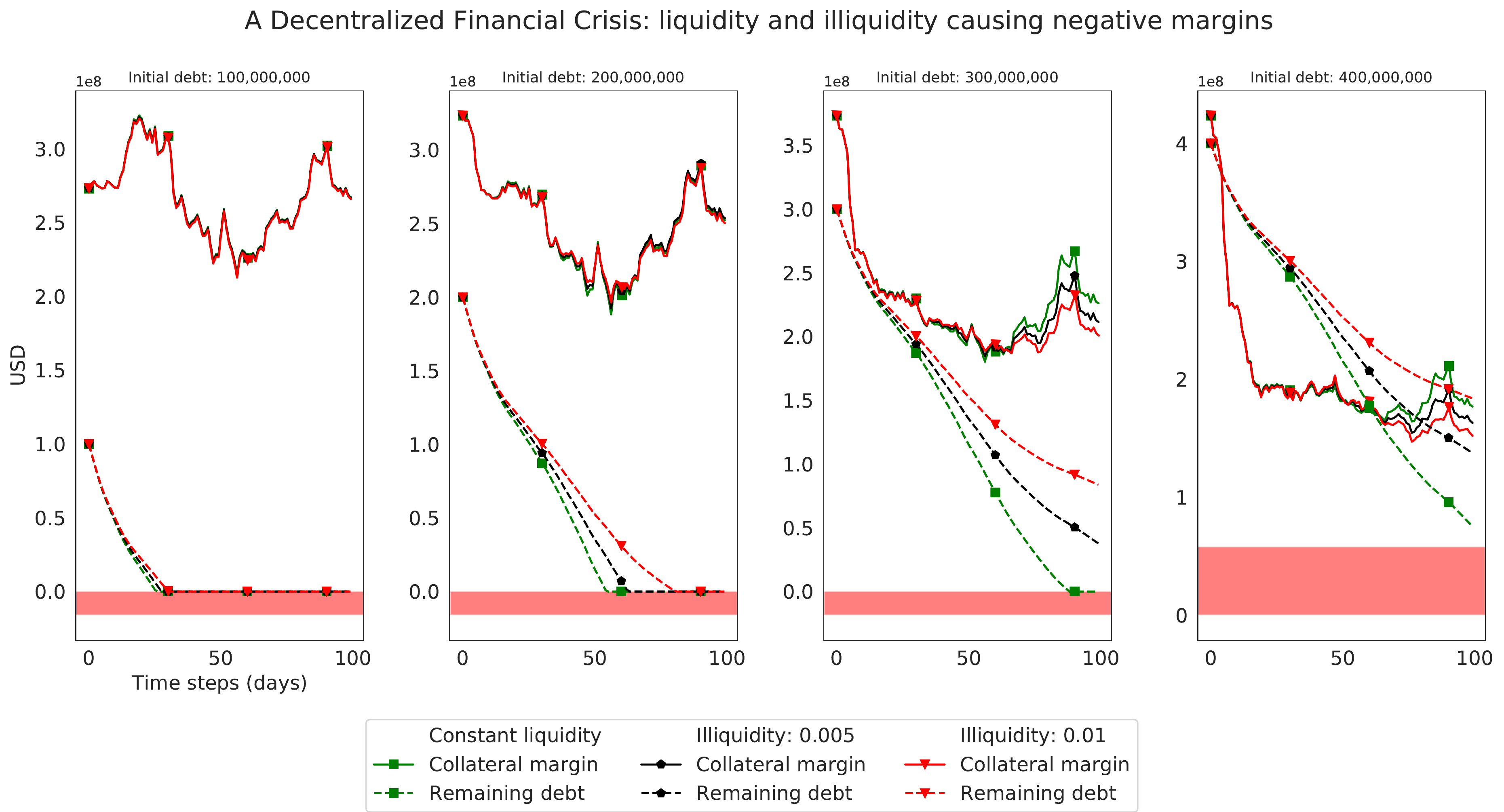}
    \caption{A DeFi lending protocol experiencing a sharp decline in the price of its collateral asset with a negatively correlated reserve asset, where the assets have a correlation of -0.9. Panels correspond to 4 different levels of system debt, with each panel showing the evolution of the collateral margin and the total debt outstanding. Each panel also shows the consequences of different liquidity parameters.}
    \label{fig:final-simulation-strong-negative}
\end{figure*}

\end{document}